\documentclass[10pt,journal,compsoc]{IEEEtran}
\usepackage{microtype}

\ifCLASSOPTIONcompsoc
  \usepackage[nocompress]{cite}
\else
  \usepackage{cite}
\fi
\ifCLASSINFOpdf
   \usepackage[pdftex]{graphicx}
   \graphicspath{{../pdf/}{../jpeg/}}
   \DeclareGraphicsExtensions{.pdf,.jpeg,.png}
\else
\fi
\usepackage{amsmath}
\newcommand{\annot}[1]{
\iftrue 
\textbf{[#1]}
\fi
}

\newcommand{\eg}{e.\,g.}
\newcommand{\ie}{i.\,e.}

\usepackage[hang,scriptsize,sf]{subfigure}
\usepackage{algorithmic}
\usepackage{textcomp}

\usepackage[ruled,vlined]{algorithm2e}

\usepackage[usenames,dvipsnames,svgnames,table]{xcolor}
\usepackage{soul}

\usepackage{hyperref}
\hypersetup{pdftex,
   pagebackref=true,
   pdfpagemode=UseNone,
   pdftitle={\@title},
   pdfauthor={\@author},
   pdfsubject={},
   pdfkeywords={},
   pageanchor=true,
   plainpages=false,       
   hypertexnames=false,    
   bookmarks=true,         
   bookmarksnumbered=false,
   bookmarksopen=true,     
   bookmarksopenlevel=3,   
   pdfpagemode=UseNone,    
   pdfstartview=Fit,       
   pdfborder={0 0 0},      
   breaklinks=true,        
   colorlinks=false,       
   nesting=true,
   linktocpage}

\begin{document}
%
\title{Molecumentary: Scalable Narrated Documentaries Using Molecular Visualization}
%
%
%
%

\author{David~Kou\v{r}il,
  Ond\v{r}ej~Strnad,
  Peter~Mindek,
  Sarkis~Halladjian,\\
  Tobias~Isenberg,
  M.~Eduard~Gr\"{o}ller,
  Ivan~Viola
  \IEEEcompsocitemizethanks{\IEEEcompsocthanksitem D. Kou\v{r}il is with TU Wien. E-mail: dvdkouril@cg.tuwien.ac.at.
    \IEEEcompsocthanksitem O. Strnad and I. Viola are with King Abdullah University of Science and Technology (KAUST), Saudi Arabia.
    \IEEEcompsocthanksitem P. Mindek is with TU Wien and Nanographics GmbH.  M. E. Gr\"{o}ller is with TU Wien and the VRVis Research Center.
  \IEEEcompsocthanksitem S. Halladjian and T. Isenberg are with Universit{\'e} Paris-Saclay, CNRS, Inria, LRI, France.}
\thanks{Manuscript version: \today.}}
\markboth{}%
{Shell \MakeLowercase{\textit{et al.}}: Bare Demo of IEEEtran.cls for Computer Society Journals}
%



\IEEEtitleabstractindextext{%
\begin{abstract}
We present a method for producing documentary-style content using real-time scientific visualization.
We produce molecumentaries, \ie, molecular documentaries featuring structural models from molecular biology. We employ scalable methods instead of the rigid traditional production pipeline.
Our method is motivated	by the rapid evolution of interactive scientific visualization, which shows great potential in science dissemination.
Without some form of explanation or guidance, however, novices and lay-persons often find it difficult to gain insights from the visualization itself.
We integrate such knowledge using the verbal channel and provide it along an engaging visual presentation.
To realize the synthesis of a molecumentary, we provide technical solutions along two major production steps: 1) preparing a story structure and 2) turning the story into a concrete narrative.
In the first step, information about the model from heterogeneous sources is compiled into a story graph.
Local knowledge is combined with remote sources to complete the story graph and enrich the final result.
In the second step, a narrative, \ie, story elements presented in sequence, is synthesized using the story graph.
We present a method for traversing the story graph and generating a virtual tour, using automated camera and visualization transitions.
Texts written by domain experts are turned into verbal representations using text-to-speech functionality and provided as a commentary.
Using the described framework we synthesize automatic fly-throughs with descriptions that mimic a manually authored documentary. Furthermore, we demonstrate a second scenario: guiding the documentary narrative by a textual input.%
\end{abstract}

}

\maketitle

\IEEEdisplaynontitleabstractindextext

%
\IEEEpeerreviewmaketitle

\IEEEraisesectionheading{\section{Introduction}\label{sec:introduction}}

%
%
%
%

\IEEEPARstart{S}{cientific} visualization has been helping researchers to make sense of their data. Visualization today also contributes to another, increasingly important, part of science: science outreach~\cite{Varner:2014:biology-science-outreach}. A growing number of researchers now focuses on communicating the current state-of-the-art of life sciences not only to students and stakeholders, but also to the general population.
Moreover, while many visualization techniques for biology have been developed, they mostly focus on transforming raw data into purely visual representations. A major issue is that, in most cases, the final image is incomprehensible to non-experts without some sort of guidance and description.
Learning is possible only at specific locations where domain-expert guidance is available, \eg, schools, museums, or science centers.

Visualization used as a tool to gain insights into scientific data only works if the user is familiar with the concepts of the particular field. Without such domain knowledge, the visual representation remains a pretty image.
A plethora of written materials exists in the life sciences (\eg, textbooks, online educational sites) with detailed information about the studied topic. In this medium, however, the visual and spatial characteristics of the matter are disconnected from the written explanations.
Consequently, a new way of learning is becoming ubiquitous and preferred by students of life sciences today~\cite{Daly:2014:microscopy-movies}. Scientific concepts are often presented using computer-generated animations and uploaded to sites such as Youtube or Vimeo. These videos communicate a topic in an engaging way by leveraging storytelling techniques developed by the animation industry over decades. A verbal narration is often an essential part that contributes to the explanatory value of such material.

Yet, pre-rendered computer animations are significantly different from interactive 3D visualizations. Mainly, a computer animation undergoes a production pipeline and often cannot easily be changed after it is published, \eg, according to new scientific findings. In contrast, an interactive 3D visualization that is rendered in real-time can provide visuals immediately on demand. Developing the visuals based on real-world data makes them flexible and ready for future extension. These aspects make 3D visualization a suitable candidate for science communication, as exemplified by its application in astronomy communication~\cite{bock:2020:openspace-new}.
The existing cases of applying visualization in science communication underline the need for incorporating explanation and guidance for public dissemination.
We pose the following research question: \emph{How can explanatory information about the function and role of individual subparts be integrated into a 3D visualization while leveraging the benefits provided by interactivity?}

\begin{figure*}
	\centering
	\setlength{\subfigcapskip}{-4ex}
	\textcolor{white}{\subfigure[\hspace{.29\linewidth}]{%
	\includegraphics[width=.32\textwidth]{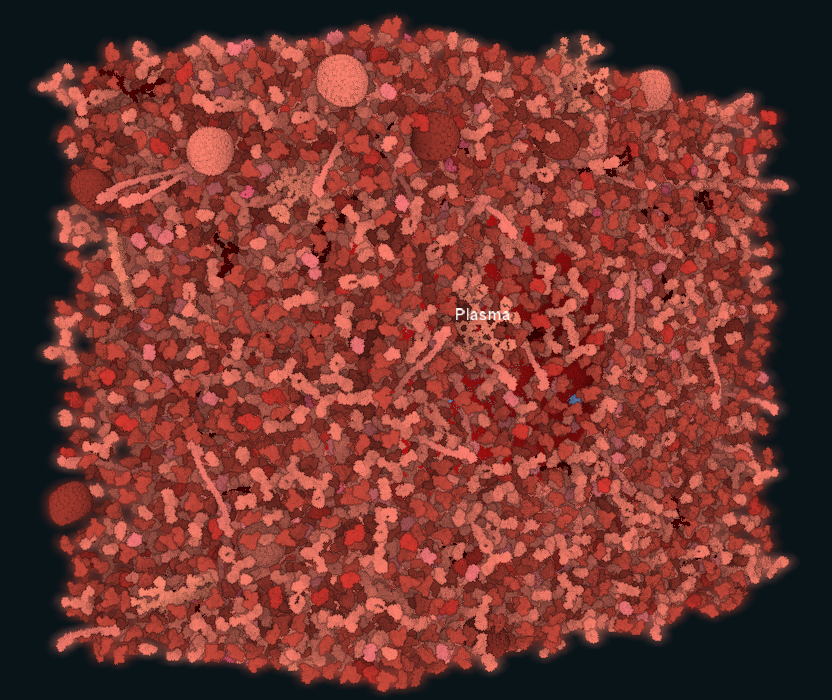}%
	\label{fig:results-hiv1}
	}}\hfill%
	\textcolor{white}{\subfigure[\hspace{.29\linewidth}]{%
	\includegraphics[width=.32\textwidth]{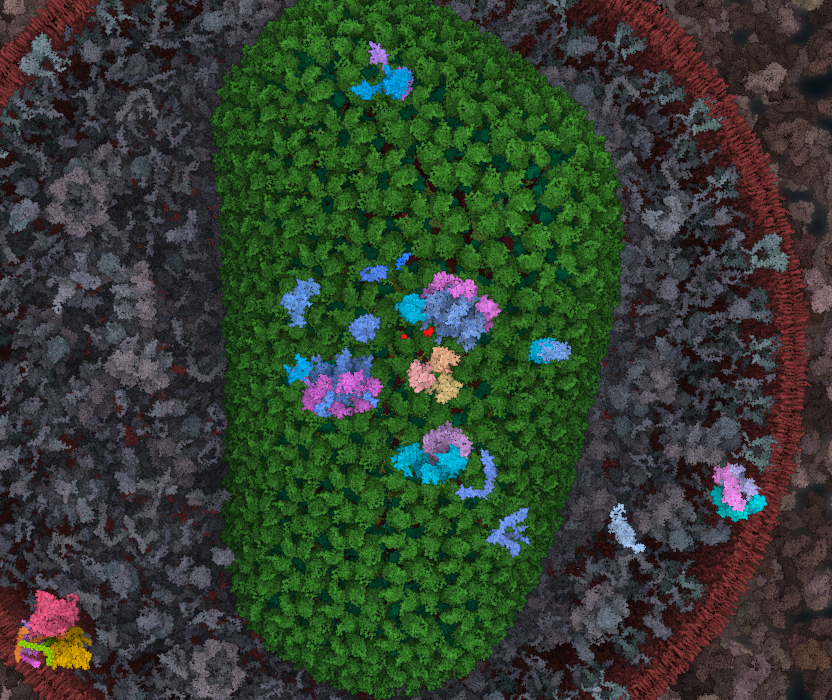}%
	\label{fig:results-hiv2}
	}}\hfill%
	\textcolor{white}{\subfigure[\hspace{.29\linewidth}]{%
	\includegraphics[width=.32\textwidth]{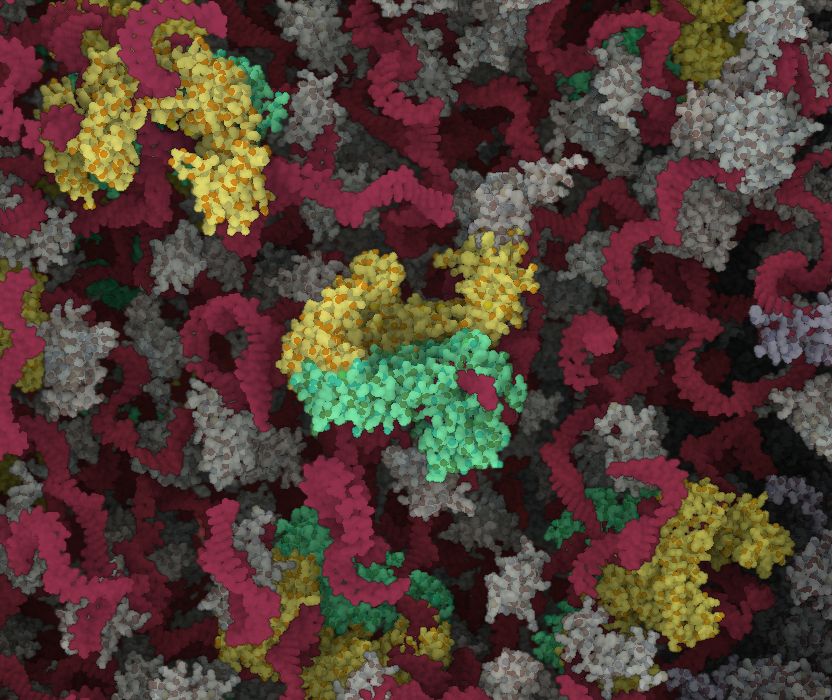}%
	\label{fig:results-hiv3}
	}}\vspace{-1ex}
        \caption{In the tour of the HIV in blood plasma model we for example visit the capsid \subref{fig:results-hiv2} which contains the genetic information of the virus. Besides the RNA, the capsid contains several important proteins, such as Reverse Transcriptase \subref{fig:results-hiv3}.}\vspace{-1ex}
	\label{fig:results-hiv}
\end{figure*}

We address this question with a method that elevates 3D scientific visualization into a scientific documentary (\eg, \autoref{fig:results-hiv}).
The explanatory information is integrated through verbal annotation using the auditory channel. We couple verbal annotations (\ie, the commentary) with an automatic fly-through of a structural model, providing visuals relevant to the commentary. The annotation communicates the roles and functions of the building blocks of the model, resulting in a virtual guided tour of the particular model. The result resembles a manually authored scientific documentary.
Our method is completely data-driven based on the structural 3D model. We describe methods for generating fly-throughs of the model, using the hierarchical organization and the functional relationships between its components.
Furthermore, we produce the verbal annotations with text-to-speech technology, which allows us to leverage content written by domain experts over many years.
This makes our method scalable and suitable for life science communication, where it is highly likely to incorporate new knowledge in the future.

To realize this novel method of using real-time scientific visualization for science communication, we contribute:
\begin{itemize}
\item the conceptual \emph{Scalable Documentary} framework which comprises real-time methods for producing scientific documentaries in a scalable and future-proof way;
\item \emph{Molecumentaries} as an exemplary application of the scalable documentary concept using multi-scale, multi-instance, and dense 3D molecular models;
\item an automated method for \emph{Story-Graph Foraging}, \ie, gathering descriptional information about the model components and constructing the story structure from these descriptions; and
\item a method for \emph{Real-Time Narrative Synthesis}, which interactively plans a traversal of the story graph, manages automatic cinematic camera animations, and ensures that a corresponding verbal commentary is provided with the visuals.
\end{itemize}

\section{Related Work}
\newcommand\sh[1]{\textcolor{Green}{#1}}

Our work relates to several areas. A molecumentary is a storytelling tool intended for the general audience. We thus start by reviewing storytelling literature. On a technical level, automatic camera control, cinematography, and the incorporation of audio are other areas that contribute to make a molecumentary and that we review individually.

\subsection{Storytelling}

One of the main purposes of visualization is to transform information (which can be extracted from data) into knowledge. Storytelling can play a key role in helping viewers to absorb this knowledge and make visualizations intelligible to a variety of audiences \cite{kosara2013, Gershon2001, Ma2012, segel-heer:2010:narrative-visualization}. Lee et al.~\cite{lee2015} discuss how storytelling should be scoped within the visualization context, and visualization research has often relied on storytelling in the past \cite{tong:2018:storytelling}.

Storytelling and visualization are thus tightly related. Storytelling communicates the visualization designer's intentions to the viewer, and visualization as a technique is a medium that allows storytelling to achieve its communication goals. Narrative visualizations combine conventions of communicative and exploratory information visualization to convey an intended story. Hullman and Diakopoulos ~\cite{Hullman2011} discussed different types of rhetorical techniques in narrative visualizations. Gratzl et al.~\cite{gratzl2016clue} presented a method for authoring visualization-based stories by integrating the stages of data exploration, story creation, and presentation. Hence, the story can be produced from the provenance information of the exploration performed by an expert user. Kwon et al.~\cite{kwon2014} proposed a web-based method for an intuitive enhancement of articles with visualizations, by linking them to relevant text segments.
Machine learning has also been employed for generating textual captions from charts~\cite{liu:2020:autocaption}.
Ren et al.~\cite{ren2017} analyzed the design space of annotations used in information visualization, and developed a tool for augmenting charts with annotations. According to their analysis, an annotation in the context of visual data-driven storytelling is characterized by two design dimensions: form and target. An annotation's purpose cross-cuts these two dimensions. In a molecumentary, text (form) annotations next to biological data items (target) enable the communication.

Storytelling methods have been applied in various areas of scientific visualization as well~\cite{Ma2012, akiba:2010:aniviz}. Wohlfart and Hauser~\cite{Wohlfart:2007} proposed a method for sto\-ry-dri\-ven presentation of volumetric datasets. Hsu et al.~\cite{Hsu2011} created multi-scale overviews of various datasets in a single image. Th\"ony et al.~\cite{thony2018} discussed various storytelling concepts, providing an overview of the design and requirements for interactive storytelling within the area of 3D geographic visualization. Lidal et al.~\cite{lidal2012} described a sketch-based storytelling technique for capturing geological interpretations during ter\-rain-ex\-plo\-ra\-tion processes. Sorger et al.~\cite{sorger2017} proposed \emph{Metamorphers}---a set of operators for authoring visual stories form molecular datasets by transforming them into comprehensible animations. In contrast to other storytelling techniques applied in scientific visualization, they took the hierarchical organization and the crowdedness of molecular datasets into account. We also focus on visualizing datasets exhibiting specific characteristics, \eg, with limited visibility and overall complexity.

Various methods have also been developed for explaining complex datasets, with possible applications in education.
Liao et al.~\cite{liao2014} visualized volume datasets by creating animations. Their se\-mi-au\-to\-ma\-tic method eases the creation of such animations by analyzing the user's interactions during an exploration of the dataset. V\'{a}zquez et al.~\cite{Vazquez:2008:interactive-anatom-edu} linked text descriptions from an electronic anatomy textbook with annotated images of 3D models, to support teaching of the anatomy. Molecumentary builds on these two approaches by producing animations enhanced by text descriptions. 
Our scalable documentaries do not rely on user interactions but on text, fetched from a various sources that describe the dataset. These descriptions are often available for biological data, but are typically hard to understand for novices without accompanying visuals. Our method thus has the potential of benefiting education and knowledge dissemination.

\subsection{Camera control}
Creating informative 3D visual narratives requires a suitable method for camera control. Approaches for automatic camera animation have been proposed for several applications. 

For multiscale environments, Van Wijk and Nuij~\cite{vanWijk:2003:zooming-and-panning} presented a theoretical framework for 2D map navigation with zooming and panning. Their concept has been used in many fields and has also been extended to 3D settings~\cite{ahmed-2005:auto-camera-graph-nav}.

Christie et al.~\cite{christie:2008:overview} presented a detailed overview of the problem to control a camera in virtual 3D environments. According to them, automated camera control is needed in common computer graphics applications, more specifically for our work, in multimodal systems. In the particular case of graphics and language (text or speech), the linguistic reference to an object dictates that the latter is not occluded, for example by using cutaways and ghosting~\cite{seligmann:1993:supporting}. In addition, spatial prepositions (\eg, in front of, left of) and dimensional adjectives (\eg, big, wide) add constraints to the camera.

The path of a moving (virtual) camera is also an important aspect of camera control. Collision avoidance in complex 3D environments, visibility of multiple targets, and smoothness of the trajectory are all challenges for the path computation~\cite{christie:2008:overview}. Salomon et al.~\cite{salomon:2003:interactive} used path planning for interactive navigation in complex 3D environments. To achieve this goal, Oskam et al.~\cite{oskam:2009:visibility} constructed a vi\-si\-bi\-li\-ty-aware roadmap graph and pre-com\-pute an estimation of the pair-wise visibility between all elements of the environment. Then, their algorithm traverses the graph and computes large, collision-free transitions in real-time.
Path planning techniques have also been proposed for digital cameras, particularly those attached to drones. Galvane et al.~\cite{galvane:2018:directing} proposed a cinematographic path planning technique. The authors relied on a vi\-si\-bi\-li\-ty-aware roadmap, similar to Oskam et al.~\cite{oskam:2009:visibility}. They generated a qualitative path (in terms of cinematographic properties) by finding the shortest path in Toric space, instead of in world space. N\"{a}geli et al.~\cite{nageli:2017:drone} applied local avoidance techniques to follow pre-designed camera paths. Both of these techniques take dynamic targets and obstacles into account.
Kn\"{o}belreiter et al.~\cite{Knoebelreiter:2014:architectural-flythrough} also proposed an automatic generation of flythroughs for architectural repositories.
Finally, Christie et al.~\cite{christie:2005:survey} provided a survey of methods of virtual camera planning.

In addition to viewpoint placement and camera path planning, cinematography addresses other issues such as shot composition, lighting design, staging (actor and scene element positioning), and editor requirements~\cite{christie:2008:overview}. Yet molecumentaries are automated documentaries, so we have no control over staging and have no editor.

\subsection{Visualization and cinematography}
Amini et al.~\cite{Amini2015} analyzed professionally designed data videos and cinematography and data design workshops. They extracted principles useful in designing comprehensible data videos. Burtnyk et al.~\cite{Burtnyk:2002:StyleCam} proposed \emph{StyleCam}---a system of camera control integrated into a 3D model viewer. Their system was designed to create highly stylized animations of 3D scenes, such as commercials or feature films. In their follow-up work, Burtnyk et al.~\cite{Burtnyk:2006:ShowMotion} introduced a method for presenting 3D models by using a dynamic camera. Their system \emph{SlowMotion} used various cinematic transitions to maximize the visual appeal of the presented model. Mindek et al.~\cite{mindek2015} developed a method for summarizing multiplayer video games, that merges views from the participating players (\ie, \emph{flock of cameras}) into a single coherent movie, visually narrating the gameplay.
Lino et~al.~\cite{lino:2010:cinematography} proposed a fully automated system that produces movies of 3D environments in real-time, with a focus on cinematographic expressiveness. By taking the 3D environment and narrative elements as input, their system computed Director Volumes with multiple associated cinematographic properties (visibility, camera angle, shot size, etc.). Then, the system edits the Director Volumes by enforcing continuity rules and computing transitions.
In our work, we use several concepts from automatic camera control and cinematography to maximize the information value of our generated visual narratives. We mainly focus on navigation within dense environments of multiscale molecular models, where not all of the common practices can be efficiently applied.

\subsection{Audio in Visualization}
A visualization can be enhanced with audio in two ways. Sonification refers to the use of non-speech audio to support visualization in communicating the intended message. Various data sonification toolkits and frameworks have been proposed, such as \emph{Porsonify}~\cite{Madhyastha1995} or \emph{Listen}~\cite{Wilson2020}.
The second way is to apply voice-overs. They could be either performed by voice actors and recorded, or synthesized. There are various surveys on the vast number of works related to text-to-speech synthesis \cite{Karpe2018, Siddhi2017}. In our method, we utilize a text-to-speech synthesizer to generate the voice narration as a key element of our automatically-generated narratives.

\section{Scalable Documentary: Overview}
\begin{figure}
\centering
\includegraphics[width=\columnwidth]{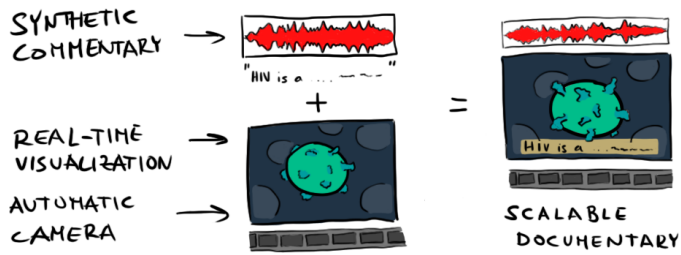}\vspace{-1ex}
\caption{The Scalable Documentary concept: We provide visuals as a real-time visualization and couple it with an automatic camera traversing the 3D scene. We augment the fly-through with a synthetic commentary that we generate on-demand, as opposed to using a pre-recorded voice-over.}
\label{fig:scalable-documentary}
\end{figure}

\begin{figure*}[t!]
\centering
\includegraphics[width=\textwidth]{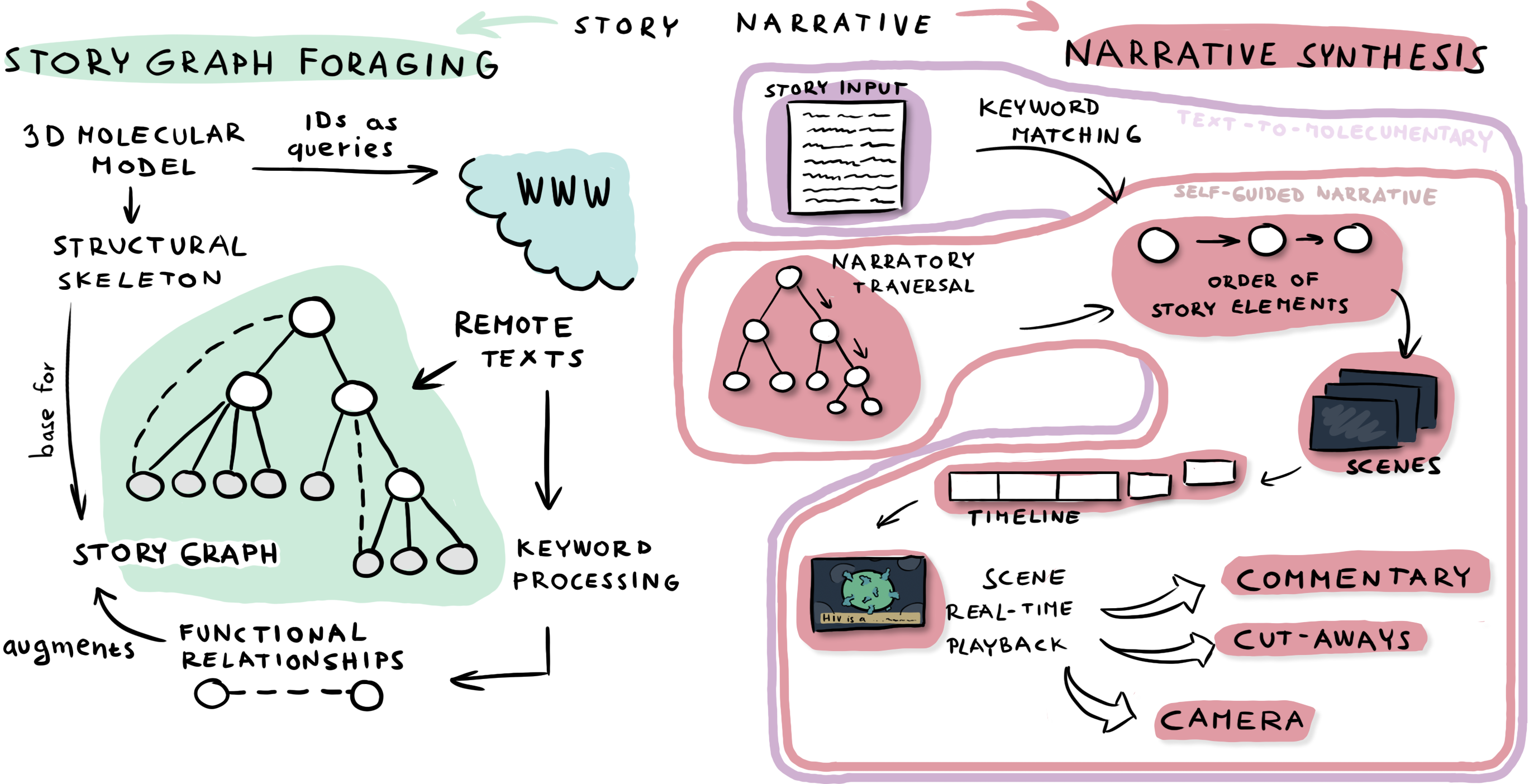}\vspace{-2ex}
\caption{Overview of the framework for generating molecumentaries. In describing the technical contributions we follow the distinction between a story (static representation of story elements) and a narrative (dynamic arrangement of these story elements). Story-graph foraging provides a scalable approach for compiling information about the model which can be used for storytelling, while the real-time narrative synthesis encompasses solutions for turning the story graph into a specific narrative at runtime.}\vspace{-1ex}
\label{fig:overview}
\end{figure*}

To address needs in science communication, we propose the concept of \emph{scalable documentaries} (\autoref{fig:scalable-documentary}), \ie, a conceptual framework in which we use interactive visualization as a medium of science communication.
We are inspired by scientific documentary movies, which explain concepts by combining computer animations and voice-over commentaries.
As the name implies, we place emphasis on scalability, \ie, the ability to adapt to future inputs.
Our framework rests on three basic components: the use of \emph{real-time visualization} instead of pre-rendered animations, an \emph{automatic camera} that procedurally traverses and showcases the 3D scene, and the coupling of visuals with a \emph{synthetic commentary}.

\textbf{Real-Time Visualization:}
Real-time visualization based on actual data, as opposed to off-line rendering, allows us to introduce a high degree of interactivity. Changes to the camera position and orientation, scene lighting, and animations are immediately reflected in the visual output.
The scene can also be dynamically modified to emphasize specific objects that are initially not visible due to occlusion.

\textbf{Automatic Camera:}
Due to our use of interactive graphics we no longer can or have to record camera movements in advance.
While it would be compelling to give users full control over what they wish to interactively explore, in complex multi-scale 3D environments such as in biology (\eg, a cell's inner structure) they may easily get overwhelmed and loose orientation.
This issue applies particularly to our target audience of lay-persons with only a limited domain knowledge.
We thus instead rely on an algorithm-con\-trolled camera and turn the visualization user into a viewer.

\textbf{Synthetic Commentary:}
A verbal commentary is an important part of a traditional documentary. The usual approach of pre-recording commentaries, however, does not fit into the concept of scalable documentaries. Flexibility is needed in this context,
new knowledge is likely to be discovered and will need to be incorporated in the future.
We thus use a procedural approach to provide the voice-over. First, we use text content written previously by domain experts. Second, we employ text-to-speech functionality to turn these textual descriptions into a verbal representation. As a result, we imitate a human commentator in a scalable way.
Furthermore, this approach is, in general, language-agnostic: Texts can be queried in any given language and we can then use an appropriate speech synthesis engine.

\smallskip

We envision our scalable documentary concept to be applicable in several scientific domains.
For the remainder of this article, however, we demonstrate it in the context of meso-scale molecular models.
This exemplary case scenario represents a situation where state-of-the-art visualization methods can produce astonishing imagery.
The visuals themselves are, however, mostly incomprehensible to people untrained in the domain.
As a proof-of-concept we produce a scalable documentary movie that integrates additional domain knowledge and provides explanation to novices.

The involved biological models are represented and visualized on a molecular level. They exhibit several specific characteristics that we need to consider in the documentary production.
First, they comprise multiple scales, exhibiting structures on scales ranging from individual atoms (approx.\ 0.1\,nm) up to a level of a whole virus (120\,nm) or even a whole cell (approx. 10~{\textmu}m).
Second, they rely on multiple instancing, \ie, the components that build up the model are present in large quantities, which also results in a dense packing. This density can lead to a cluttered image and we thus need to incorporate cut-aways to show the inner composition of the model.
Moreover, the models are constructed through a specific technique~\cite{cellpack, cellPACK_HIV_Johnson2014} with an inherent labeling, \ie, identifiers of their components.
We use these universal object identifiers in the documentary synthesis to reach an even greater level of scalability.

Before we explain the technical details of how we generate a molecumentary, we define two terms that are often used interchangeably---a \emph{story} and a \emph{narrative}. For the purpose of our method, we differentiate between these two.\footnote{We base our terminology on Olaf Bryan Wielk's, a storytelling theoretician:
https://www.beemgee.com/blog/story-vs-narrative/}
We consider a \emph{story} to be the overall architecture of story elements, \eg, events, actors, and their relationships. In contrast, we regard a \emph{narrative} as a sequence of these story elements presented in a certain order. Different narratives of the same story can be built by changing the order of story elements.
We organize the technical description of our framework along this distinction between a story and a narrative, as illustrated in~\autoref{fig:overview}.

In \autoref{sec:semantics-foraging} we explain to organize a story in a data structure, which we call \emph{story graph}.
The story graph holds all the model elements, their relationships, and meta-information regarding the biological model.
Creating such a story structure manually is tedious and in the context of a molecumentary would require the involvement of a domain expert. We present \emph{Story Graph Foraging} as an automatic method for constructing the story graph. 
In Story Graph Foraging we fetch descriptions about the model components from both local and remote sources, and then extract relations between the components from these textual descriptions.

In \autoref{sec:docu-synthesis}, we describe how we generate the actual molecumentary.
We use the story graph to generate an on-the-fly narrative, \ie, we build a sequence of story elements that will be featured in the molecumentary. Furthermore, in the narrative generation we use the descriptions stored in each story element to synthesize an on-demand commentary, using text-to-speech functionality. With automated camera animations and occlusion management we execute scenes that communicate the subcomponents of the model.
We determine the order of the model elements shown in the documentary in one of two ways. In the first case, the documentary is self-guided, \ie, an algorithmic approach determines in which sequence the hierarchical structure of the model is explored. In the second case, which we call \emph{text-to-molecumentary}, we generate visuals that follow a storyline supplied as written-text input.
Moreover, the visuals can react to changes in the text directly, so the whole system can be used in real-time. The user can textually compose the story and immediately see its impact.

Our proposed concept facilitates scalable science communication. By automatizing a large portion of the scientific movie production pipeline, we are able to immediately reflect new knowledge, \eg, new research results, into the science communication medium such as scientific movie, interactive learning tool, or museum installation.
We note, however, while we do deal with the terms story and narrative, that we do not attempt to solve the problem of generative storytelling. We rely on texts written by various writers, but essentially consider these texts as ``black boxes.'' We do not extract meaning and do not aim at producing a story that is creative and/or stylistically correct.

\section{Story-Graph Foraging}
\label{sec:semantics-foraging}
At the core of our method lies the \emph{Story Graph}, which contains data needed to build stories about a biological model.
The story graph is composed of \emph{type nodes} and \emph{relationships edges}.
Each of the nodes represents a type of a biological structure featured in the model and contains a set of descriptions detailing its role.
More than one edge is allowed between two nodes, which turns the story graph a multigraph.
The edges represent relationships between the structural elements. These relationships can have several types. In our work, we specifically recognize two cases: structural relationships and functional relationships. Structural relationships represent spatial and hierarchical relations of the subcomponents (\eg, \emph{blood plasma contains the protein Albumin}). Functional relationships relate structures that are involved in a certain biological function, \ie, they interact or are related.
Based on the two edge types, the story graph can be decomposed into a directed acyclic graph, which models the structural relationships (we later refer to this as the ``skeleton'' of the story graph), and a general multigraph, which contains the functional relationships.

\begin{figure}
\centering
\includegraphics[width=\columnwidth]{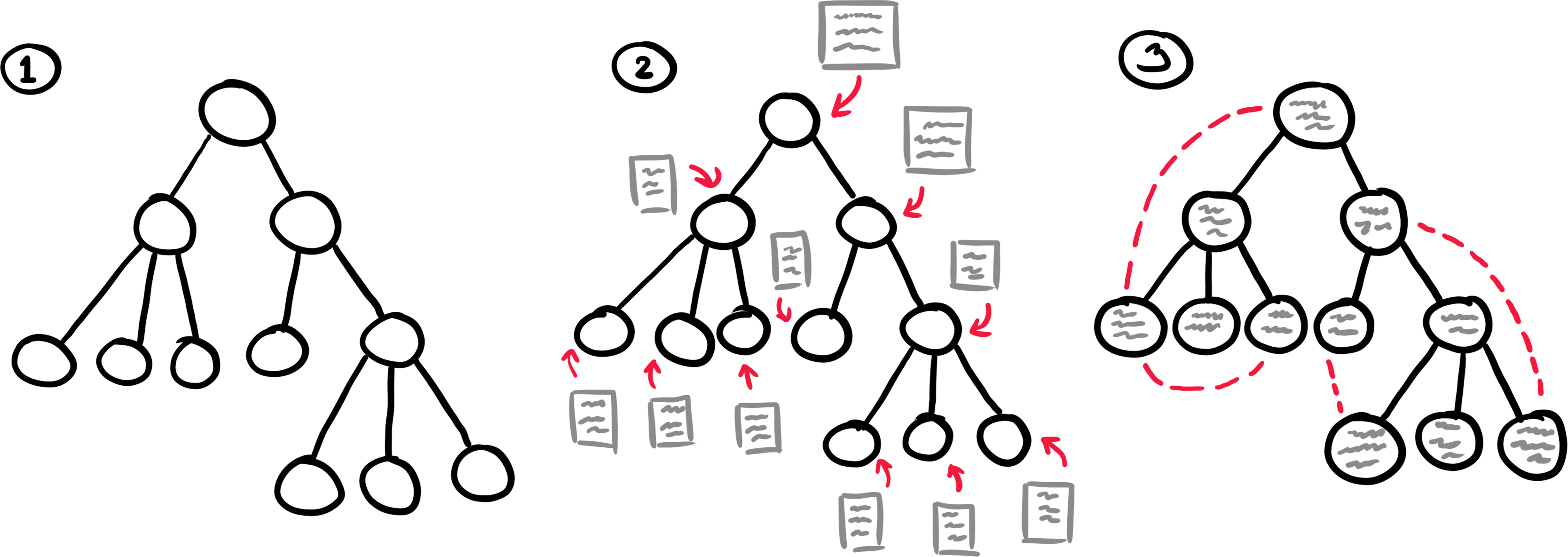}\vspace{-1ex}
\caption{The three steps in story graph foraging. First, we build the structural skeleton. Second, we associate textual descriptions with individual nodes. Finally, we add functional relationships.}\vspace{-1ex}
\label{fig:foraging-steps}
\end{figure}

In the rest of this section, we describe our method for building the story graph by \emph{foraging}.
We use the term foraging rather than construction to express the flexibility and liveliness of this process. The story graph is not only constructed once with a specific, correct result as a goal. It rather is a continuous process that can achieve different results, depending on the case and situation. This reflects the volatility of the subject matter, with new knowledge coming in, new repositories becoming available, and the large number of stories that can be told in this context.
We perform story graph foraging in several steps (see~\autoref{fig:foraging-steps}), each improving the resulting generated narrative.

\subsection{Step 1: Structural Skeleton Foraging}
In the first step, we organize the structural elements of the model into a basic skeleton of a story graph.
We establish the structural relationships between biological structures which define the structural organization of the model.
For example, for proteins building up a certain higher-level composite object, then we consider these two connected via a parent-child relationship. Such relationship is then represented by a structural relationship edge in the story graph.

We create a node for every molecular type in the model. Using the structural model, we first obtain the base skeleton for the story graph and use it to build a tree skeleton from its hierarchical organization. We then represent higher-level structures (parent objects) by their own nodes in the story graph.
The names of individual components are the only descriptive information that we can relay to the viewer at this point. They can be shown, \eg, as textual labels.
Furthermore, a simple narrative can be synthesized using even with a story graph just containing this structural ``skeleton.''
However, the output would be rather rudimentary, essentially communicating the structural organization of the biological model (\eg, ``Structure X contains components Y, Z, and W. Let us look at Y first.'').

\subsection{Step 2: Type Node Descriptions Foraging}
\label{subsec:node-descriptions-foraging}
We can improve the resulting narrative by incorporating descriptions about the individual structure types, which explain the role of the associated structure in the biological model. Gathering these descriptions represents the second step of story graph foraging. At the lowest level, the textual labels from the structural skeleton can be extended with additional or alternative \emph{names} of the structure, such as some form of identification (\eg, PDBID in case of proteins). While this minimal annotation facilitates identifying the structure and distinguishing it from others, it does not provide any insight into its function. A higher-level description of the function can thus be established by expanding this annotation to the level of \emph{one short sentence} or, of course, to \emph{longer descriptions} with more detailed explanations. In both cases, the expressiveness and possibility of clearly communicating the message largely depends on the skill of the writer.

There are several options for getting these descriptions.
First, some descriptive texts can be manually written and \textbf{supplied locally} along with the structural model. We present these text snippets with the highest priority since they are specifically created to describe the given structure. However, they might only express one level of detail and are not scalable since they have to be prepared for every element.
In case no such information is provided, we thus use an alternative way of gathering descriptions. We take the standardized names of biological structures (\ie, Albumin) as keywords for searching in \textbf{remote, online repositories}.

Publicly accessible databases (such as Wikipedia) contain a large amount of interesting and relevant information written by domain experts.
We take advantage of web APIs and use the name of the queried structure as a search keyword and fetch the structural description as a response.
We target short, high-level descriptions which describe the searched term in a few sentences. In our case, we use so-called extracts available in responses from the Wikipedia API which typically contain the first paragraph of the Wikipedia page for the described topic.
This process can be done in real-time and we do not need to pre-fetch the descriptions for the whole model before the narration starts, saving memory.
One of the biggest benefits of the real-time fetching construction is that we can scale this approach to any size of the model, provided that the model is reasonably annotated.
Also, by fetching the data online in multi-lingual databases such as Wikipedia, we can query information in several languages. Furthermore, these days very powerful translation engines are becoming available and their APIs can be used as a component in the descriptions foraging pipeline.
The drawback of this approach is that if the element is annotated by a generally known word that is typically used with different meanings in several domains (\eg, ``plasma'') the results may not be relevant. We accept this trade-off as it is easier to modify a label to be more specific than to write a reasonable paragraph of text.
A label can also contain a name for which querying does not produce
text from the remote source. In this case we fallback to the structural commentary we explain in \autoref{sec:structural-commentary}.

When descriptive texts are incorporated in the narrative synthesis, the result is a much more natural sounding documentary. The virtual narrator provides explanations to the viewer and the viewer learns about the structures visible on the screen and their functionalities.

\subsection{Step 3: Functional Relationship Edges Foraging}

\begin{figure}
\centering
\includegraphics[width=\columnwidth]{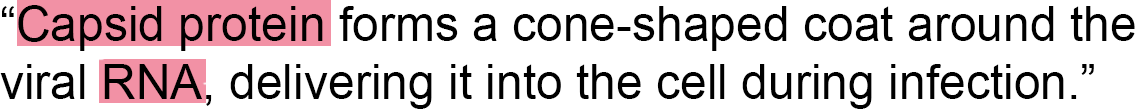}\vspace{-1ex}
\caption{A sample textual description in which a functional relationship has been extracted. Through keyword detection the fact that the capsid protein forms a structure protecting the RNA is established. Such a functional relationship is added to the story graph as an edge.}\vspace{-1ex}
\label{fig:func-relationships}
\end{figure}

So far the order of explanation is only driven by the structural organization.
To relate structures that are associated not because of their proximity in the hierarchy but rather because how they interact, we need to add functional relationships to the story graph---the final step of story graph foraging.

We could establish these functional links by using data about the metabolic exchanges between structural components of the modeled organism, \ie, the metabolic pathways. Integrating such data, however, would require the intervention from a domain expert.
We thus use another---text-based---approach to extract functional relationships, as illustrated in~\autoref{fig:func-relationships}.
We get the names of all substructures in the model from the story graph skeleton and accumulate them into a keywords list.
We then process the user-authored or downloaded texts from \autoref{subsec:node-descriptions-foraging}, split them into sentences, and search these for keywords they may contain. 
When we detect any keyword in a sentence, we establish functional relationship edges between structures associated with these keywords.
We then consider these functional neighbors when the story graph traversal method decides on which nodes shall be covered in the synthesized narrative.

\section{Narrative Synthesis}
\label{sec:docu-synthesis}
After we created the story graph with both structural and functional information, we prepare the story for the narrative synthesis.
Below we first describe the general approach for producing a specific narrative, \ie, the story elements presented in a sequence.
Next, we demonstrate two scenarios of molecumentary synthesis. In one scenario we decide what is shown solely based on our story graph traversal algorithm. In the other scenario we use an human-authored, textual narrative and employ our molecumentary synthesis process to produce accompanying visuals.

\subsection{Timeline and Scenes}
We represent a specific narrative in a \emph{timeline} data structure.
The timeline is composed of a sequence of \emph{scenes}, where each scene contains both visual and audio pieces of individual parts of the molecumentary.
The timeline works as a queue---scenes are added (pushed) to the back and removed (popped) from the front, implementing the first-in first-out approach.

\begin{figure}
	\centering
	\setlength{\subfigcapskip}{-3.5ex}%
	\textcolor{white}{\subfigure[\hspace{0.93\linewidth}]{%
	\includegraphics[width=\linewidth]{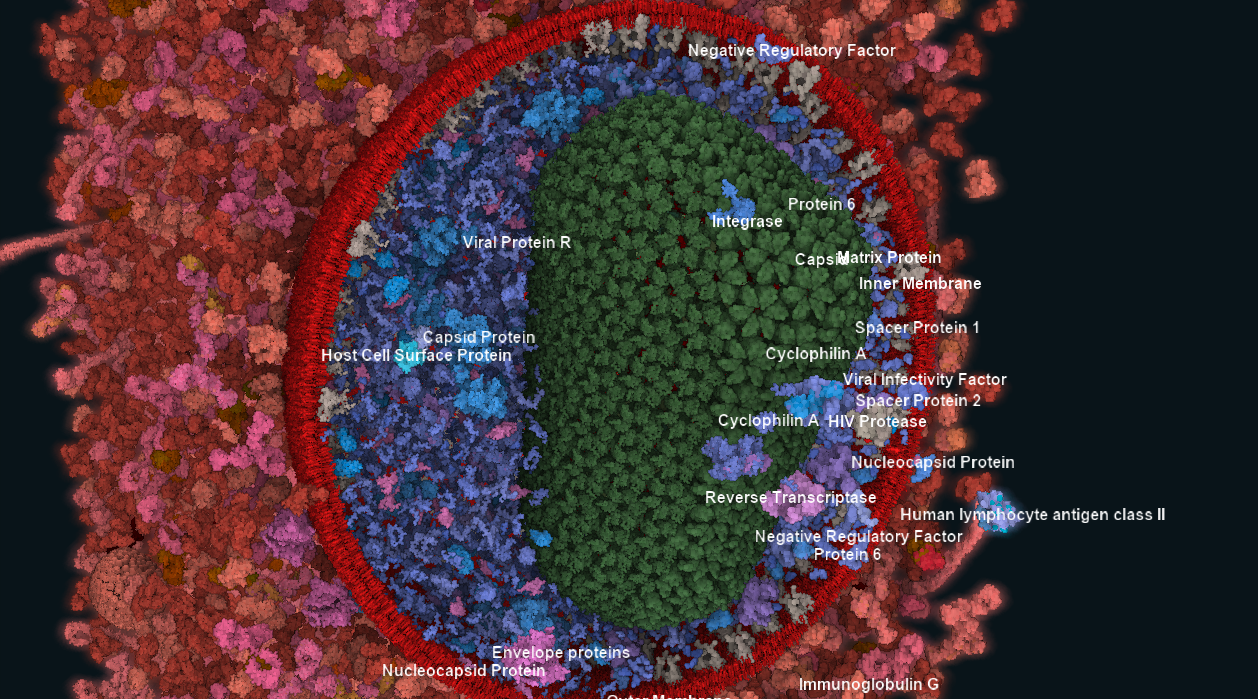}%
	\label{fig:scenes-overview}
}}\\
	\textcolor{white}{\subfigure[\hspace{0.93\linewidth}]{%
        \includegraphics[width=\linewidth]{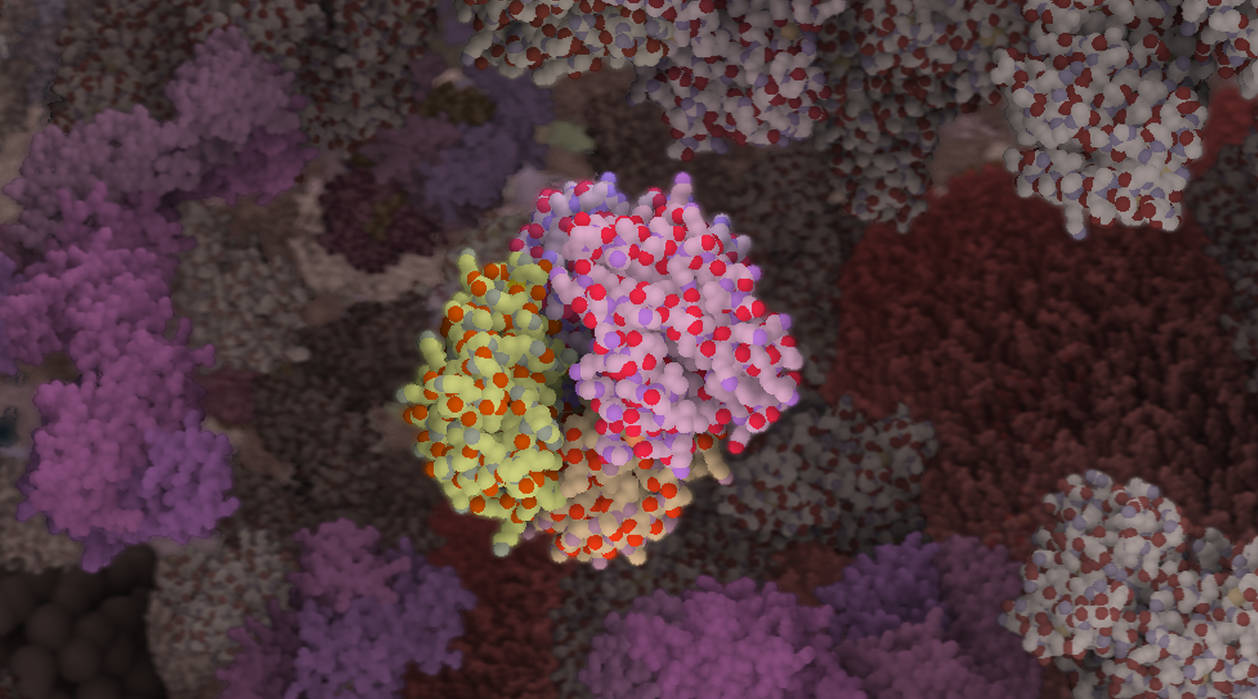}%
	\label{fig:scenes-focus}
}}\vspace{-1ex}
	\caption{Overview scene \subref{fig:scenes-overview} communicates the composition of an object while a focus scene \subref{fig:scenes-focus} describes its function. Transition scene type is defined to switch focus between different objects and connect the other two types of scenes in the narrative.}\vspace{-2ex}
	\label{fig:scene-types}
\end{figure}

We use scenes of three types: focus, overview, and transition.
A \emph{focus scene}  (\autoref{fig:scenes-focus}) is the central building block of our narrative: it shows details about one structure type. We move the camera to close in on the selected instance, then use subtle rotation animations to provide parallax, and give a detailed description of the function and responsibility of the focused object inside the modeled organism.

The issue with only using focus scenes in the molecumentary is that the object in focus is shown as a whole and the viewer does not get a good idea of its internal composition. This is particularly problematic for composite objects in the hierarchical model.
Therefore, we incorporate a second scene type: overview scenes.
An \emph{overview scene} (\autoref{fig:scenes-overview}) shows all building blocks of a certain model part. The aim is to communicate the structural composition of the object. We adjust the cut-away settings of the visualization to showcase the building blocks. We highlight representative instances for each subcomponent and place the camera at such a position that shows all of them. The accompanying commentary verbalizes these components and establishes their relationships with the current object in focus.

To be able to meaningfully switch between the various focus and overview scenes, we need animated transitions that communicate a shift of emphasis.
We use \emph{transition scenes} bridging the other two types of scenes. They mostly function as connective material between the individual overview and focus scenes and provide context for a fly-through. Transition scenes usually contain significant camera transitions and changes in the visualization. The verbal commentary in these scenes then provides additional guidance and comments the transitions that happen.

We now describe the three processes---camera animation, occlusion management, and voiceover---that we used in implementing the scenes in the molecumentary synthesis.

\subsubsection{Camera Animation}

\begin{figure}
	\centering%
	\setlength{\subfigcapskip}{-3.5ex}%
	\textcolor{white}{\subfigure[\hspace{0.93\linewidth}]{%
	\includegraphics[width=\linewidth]{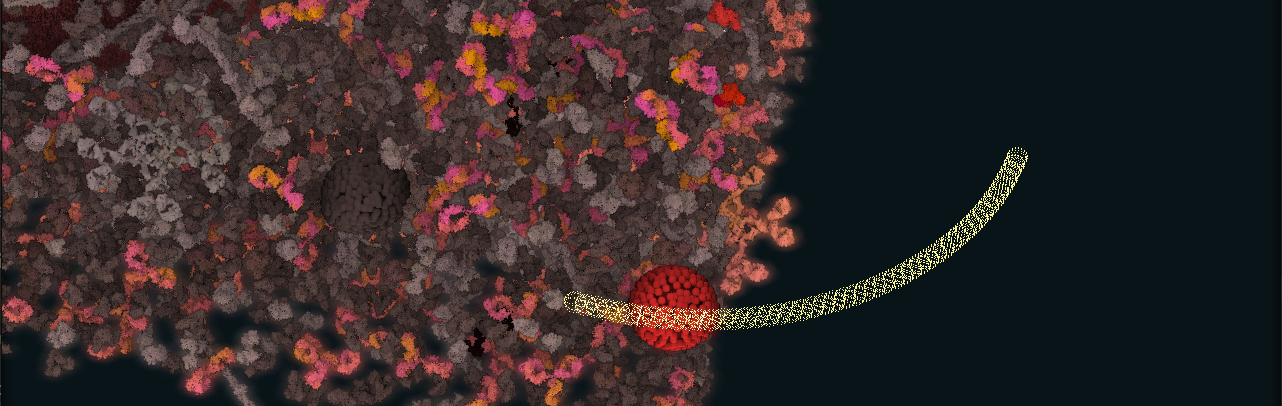}%
	\label{fig:camera-orbiting}}}\\
	\textcolor{white}{\subfigure[\hspace{0.93\linewidth}]{%
	\includegraphics[width=\linewidth]{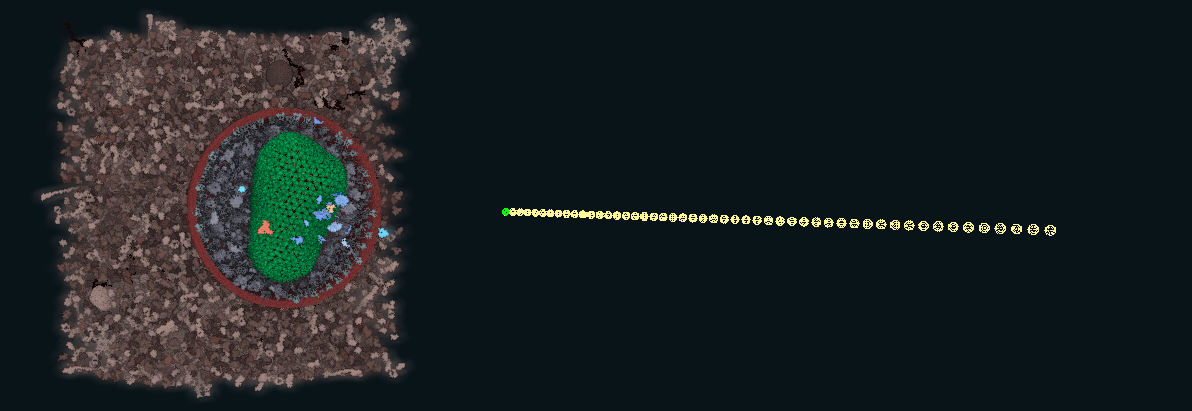}%
	\label{fig:camera-straight-flying}}}\\
	\textcolor{white}{\subfigure[\hspace{0.93\linewidth}]{%
	\includegraphics[width=\linewidth]{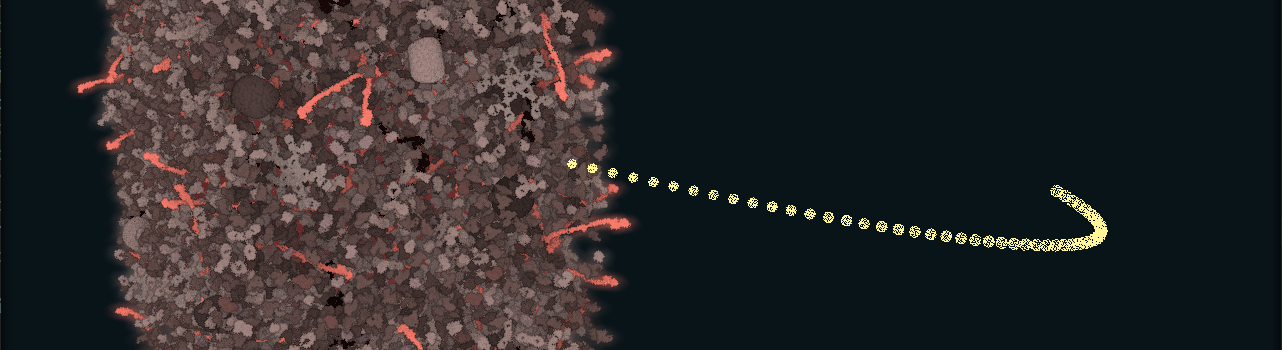}%
	\label{fig:camera-bezier}}}\vspace{-1ex}
        \caption{Illustration of the three camera animation types used in a molecumentary synthesis: anchored orbiting~\subref{fig:camera-orbiting}, direct flying~\subref{fig:camera-straight-flying}, and curved transition~\subref{fig:camera-bezier}.}\vspace{-1ex}
	\label{fig:camera-anim}
\end{figure}

Camera movement plays an important role in conveying the multi-scale model, along with its many subcomponents.
We primarily use three movement types in producing the molecumentary. These are \emph{anchored orbiting}, \emph{direct flying}, and \emph{curved transition}, illustrated in~\autoref{fig:camera-anim}.
Many more camera movements can be incorporated and developed for future applications. Here, we describe our basic camera language sufficient to be used in our prototypical implementation.

To be able to generate the camera animations we need to know the position and shape of the structures in focus in each scene.
We use a bounding sphere as an approximation of the target object shape and size. A bounding sphere can be quickly computed in real-time and in many cases in our application scenario (molecular models) it approximates the shape sufficiently well.
The camera animations are defined between targets that are specified by two attributes each: world position and a radius of the bounding sphere.

\emph{Anchored orbiting} refers to a slow movement of the camera rotating around a specific object instance while keeping the camera oriented toward the center of the instance.
Anchored orbiting achieves two goals: it provides 3D motion parallax and gives an impression of the local neighborhood. It thus contextualizes the focused instance in 3D space and shows neighboring structures.
We use anchored orbiting in focus and overview scenes.
The orbiting direction (clockwise or counterclockwise) is decided randomly in each scene.

For a continuous narrative, however, we also need to transition between two focus instances for which we use \emph{direct flying}. We animate the camera along a straight line, with its orientation fixed.
This movement type is suitable for cases where the two instances (initial and target) are visible from the initial camera viewpoint. If the target position is outside of the viewing frustum, direct flying can be suboptimal in communicating the spatial relation between the two objects.

Therefore, we introduce the third movement type: \emph{curved path animation}. In this animation type we zoom the camera slightly out of the initial focus position, providing context of its surroundings, and then travel toward the target focus position on a curved path. We use a normal B{\'e}zier curve defined by three points, but any curve type can be used.

\subsubsection{Occlusion Management}

\begin{figure}
    \centering
    \includegraphics[width=\linewidth]{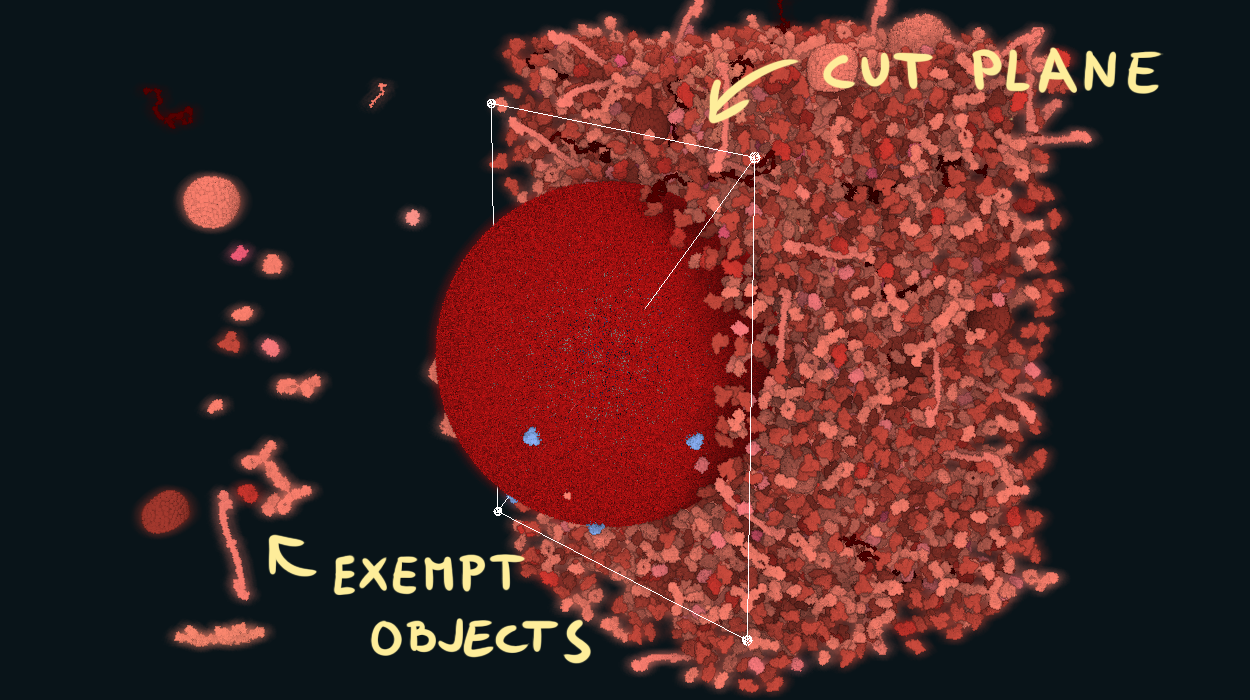}\vspace{-1ex}
  \caption{Traveling cutting plane: We remove all objects---except a selected subset---that lie between the cutting plane and the camera position to reveal inside components of the model. We implemented the cutting based on the world-space position of the scene objects. We determine the objects that are exempt from the cutting based on the scene type and the type of object that is currently shown in the molecumentary.}\vspace{-1.25ex}
  \label{fig:occl-man}
\end{figure}

Biological models are densely packed with molecules, which results in occlusion of most of the interesting structures, \eg, in the inside of a virus.
Occlusion management is required to showcase all relevant parts of the model properly.

We employ a \emph{traveling cutting plane} approach (see~\autoref{fig:occl-man}). We define a cutting plane in the scene and do not render any object that lies between the cutting plane and the camera.
We exclude, however, certain instances (or types) from being cut away.
This allows us to highlight the selected objects as well as convey the impression of the absolute number of these objects in the model.
The cutting plane \emph{travels}, \ie, we animate it and the set of objects we always show throughout the molecumentary to successively reveal objects that are being verbally described.
We perform the animated transitions in the \emph{transition scenes}.
We then determine the objects exempt from removal based on the type of the scene that follows the transition.

For a \emph{focus scene} we shift attention to one (sub-)structure type. To emphasize this focus type, we exempt all its instances from being cut for the duration of the scene to communicate their number in the model.
We then re-position the cutting plane to the center of a selected representative instance. We select the instance closest to the camera as the representative and orient the cutting plane to be parallel to the viewing plane at the moment the object comes into focus. We thus set the normal vector of the plane to be the same as the camera's initial back vector.
In an \emph{overview scene} we communicate inner composition of a structure. In the transition scene that leads up to it we thus create a view that shows the inside. We do so by fetching the structural components (child nodes) of the focus structure and, for each of these child nodes, pick a representative instance and exempt it from the cutting. We then place the cutting plane at the position of the representative that is furthest away from the camera such that none of the representatives is occluded by instances kept in the scene.

We purposefully used the traveling cutting plane as a world-space technique that culls instances, rather than image blending effects. The fading in and out of alpha blending resembles a ``cut'' in movie making, which would make it less apparent that changes in the scene communicate an opening up of the model, as opposed to a change of the scene altogether.
Also, while we incorporate only one cutting plane in our design,
we envision that using multiple planes would be possible. However, keeping track of the cutting planes to ensure that an object selected in the future will be visible, in our opinion, outweighs the potential benefits.

\subsubsection{Verbal Commentary}
We realize the verbal commentary with text-to-speech tools. We generate the commentary in textual form and later synthesize audio using an artificial voice.
We employ three types of commentaries: structural, descriptional, and navigational.

We use \emph{structural commentary}\label{sec:structural-commentary} in overview scenes to describe the structural composition of certain composite objects. In structural commentary we explain which sub-objects make up the described object. An example of structural commentary is \emph{``Blood plasma consists of Hemoglobin and Heparin and others.''}
We construct the commentary procedurally based on the hierarchical object composition. 
To make the generation process scalable, we use sentence templates. Because the structural commentary relies on a hierarchy, the sentences typically contain word formations such as ``consists of \dots,'' ``belongs to \dots,'' and similar. We further define variables that can be used in the templates. We replace these variables in real-time by respective values based on the current story graph traversal. The variables are: \emph{\$name}, \emph{\$siblings}, \emph{\$children}, \emph{\$parent}, \emph{\$previous}.
\emph{\$name} denotes the element on which the story currently focuses. The variables \emph{\$siblings}, \emph{\$children}, \emph{\$parent} contain hierarchical information related to the current node \emph{\$name}.
The variable \emph{\$previous} points to the node which has been in focus just before \emph{\$name}.
In the above example, the template we used was \emph{``\$name consists of \$children''}.
In large hierarchies, \emph{\$children} and \emph{\$siblings} can contain tens or hundreds of nodes---too many to list all in the commentary.
Therefore, we randomly select a subset of them (we use three) to keep the sentence short. Also, the commentary likely changes in case the virtual tour returns back to the current node and the structural commentary is used again.

We employ a \emph{descriptive commentary} in focus scenes.
It provides the explanatory information about the individual components of the model. We use the previously described contents (\autoref{subsec:node-descriptions-foraging}) of the story graph nodes to synthesize its text to describe the objects' functions and significance in the model. We use pre-defined texts with a higher priority than ones fetched from online sources.
We currently consider these texts as black boxes, so their expressiveness depends on their authors and we use them as is.

We use a \emph{navigational commentary} in transition scenes.
Its purpose is to contextualize what happens in the transition scene and connect the overall narration.
We synthesize the sentences using the same templating as we used in the structural commentary, but with a different set of templates (\eg, \emph{``After focusing on {\$previous} we can see {\$name}.''}), from which we select random entries.

We also display textual labels in the scene~\cite{kouril:2020:hyperlabels} to connect the verbal narration with the shown structures and to help viewers to differentiate the mentioned objects. We dynamically place labels that name the structures on those representative instances that are relevant to the current scene.

\subsection{Self-Guided Narrative}
Given the general concept for molecumentary synthesis, we now present two variants of this scalable documentary application. Here, we first showcase a self-guided molecumentary, \ie, we do not use input that would inform the narrative to be shown.
Instead, we automatically create the flythrough based on the organization of the model and a specific ``narratory traversal'' story graph exploration algorithm.

\subsubsection{Narratory Traversal}
In deciding the order of story nodes in the documentary, we deal with traversing the story graph structure. The story graph represents the hierarchical organization of the model and we wish to communicate this organization to the viewer.
In the context of the molecumentary synthesis we aim to replicate the look and feeling of a scientific movie. For such a purpose, the traditional methods of traversing a tree or a graph data structure do not provide the desirable engaging results. The usual algorithmic traversal approaches result in a mechanical exploration and typically violate our requirement for the final documentary to be engaging.

We thus propose the more captivating strategy of \emph{narratory traversal}, in which we step through the graph not with the goal of systematically visiting every node, but to showcase the 3D hierarchical structure represented by the graph. Naturally, a multitude of ways exist to meet such goals. Here we describe a method that uses two interconnected data structures: \emph{the traversal stack} and \emph{the options pool}.
A stack is data structure often used for exploring trees and graphs, and we use it to contain the nodes of the story graph.
The pool contains the options for next objects to feature in the documentary.
At any time, the top of the stack signifies the current node and, therefore, a level in the hierarchy.
The pool structure then contains all the options (\ie, nodes) that we can access directly from the current node; \ie,
(a) parent, (b) children, or (c) functionally related nodes. These nodes represent potential next targets, and we recompute the pool any time a node is pushed to or popped from the stack.

We use a stochastic approach to pick the next targets from the pool, as detailed in \autoref{alg:selection}.
Our defining criterion is whether the potential next node has been previously shown, and if so then when. We specifically use the time of last visit to ensure that we can continuously traverse the whole model if the molecumentary is left to run for longer periods.
The $Priority$ function models a priority distribution among the nodes, and we define it as
\begin{equation}
Priority(n) = \begin{cases}
P_{lower} & \text{$n$ is a leaf node} \\
P_{higher} & \text{$n$ is an inner node}
\end{cases}\:\:.
\end{equation}

\SetAlgoSkip{}
\begin{algorithm}[t]
\footnotesize%
\SetAlgoLined
\tcp{options from the pool}
$\textbf{var}~options$\;
\tcp{times of last visit}
$\textbf{var}~visitedTimes$\;
$min = getMinimumValue(visitedTimes)$\;
\ForEach{$option \in options$}{
    \If{visitedTimes[option] == min}{
        $candidates.add(option)$\;
    }
}
\ForEach{$c \in candidates$}{
    $priority = Priority(c)$\;
    $priorityRange += priority$\;
}
$rand = random(0, priorityRange)$\;
$prioSum = 0$\;
\ForEach{$c \in candidates$}{
    $priority = Priority(c)$\;
    $valA = prioSum$; $valB = prioSum + priority$\;
    $prioSum += priority$\;
    \If{$valA < rand <= prioB$}{
        $next = c$\;
        $break$\;
    }
}
$visitedTimes[next] = currentTime$\;
\Return{$next$}\;
\caption{\mbox{~Next story node selection}}\label{alg:selection}
\end{algorithm}

It is also possible to incorporate manual input in the priority function, \eg, based on expert opinion for significance of a  specific subset of structures in the model.

\subsubsection{Timeline Building and Playback}
The step-wise procedure for determining which nodes will be featured in the tour, however, is not yet sufficient. To produce a molecumentary we still need to turn this node sequence into a sequence of scenes that we can place onto the timeline.
When a node (\ie, object type) is selected to be shown, we thus first add a transition scene from the current object in focus to the new one is put into the timeline, followed by a new focus scene for the newly selected node instance.
In addition, if the selected node is a composite object (\ie, an inner node in the structural skeleton of the story graph) we perform a ``diving into'' operation: We push the node onto the traversal stack, which leads to the pool of options being recomputed. We then generate an overview scene to convey the composition of this object, after we first added a transition scenes to introduce the coming composition explanation.
We detail the procedure in \autoref{alg:self-guided}.

\begin{algorithm}[t]
\footnotesize
\SetAlgoLined
 $lastScene = timeline.last$\;
 \eIf{$lastScene.type == overview$}{
     $transitionOverviewToFocus(current, next)$\;
 }{
     $transitionSiblings(current, next)$\;
 }
 $focus(next)$\;
 \If{$isLeaf(next) == false$}{
    $pushToStack(next)$\;
    $transitionFocusToOverview(next)$\;
    $pushOverview(next)$\;
 }
\caption{\mbox{~Scene generation (self-guided narrative)}}\label{alg:self-guided}
\end{algorithm}

\subsection{Text-To-Molecumentary}
Often there already exists a description of a particular model that describes the important parts and their functional behavior. Our second synthesis variant thus uses a story in a textual form as input to generate the molecumentary.

We parse the input text by sentences. 
In each sentence we search for the names of structures in the model and fetch the corresponding story graph node if we have a match.
To prevent keywords that are frequently mentioned in the input text from being focused on and shown multiple times, we use every detected keyword only once during the whole story.
Furthermore, we want to avoid many focus shifts within a short period of time.
If multiple keywords are detected in a sentence, we thus use only the first keyword that has not yet been excluded as a story element.

In a second step we convert the found story graph nodes (\ie, structural types) into a series of scenes, similarly as we did it in the self-guided version. We then push these scenes to the timeline, which we later play in the same manner as explained before.
The approach
for generating the scenes also takes into account the hierarchical relationship between what was shown before and what shall be shown next in the molecumentary.
Since the narrative in the input text can express arbitrary jumps through the hierarchy, the resulting scenes no longer communicate a node-by-node traversal of the story graph.
To clearly communicate the hierarchical and encapsulation relationships, we could inject scenes showing also elements intermediate between the current and next target. However, we choose not to do so and transition directly to the next detected node because additional scenes would disrupt the narrative and cause undesired pauses in the synthetic voice-over. In our tests this worked without problems, provided that the input text was of sufficient quality.
We summarize the approach in \autoref{alg:text-to-molecumentary}.

\begin{algorithm}[t]
\footnotesize
\SetAlgoLined
\tcp{list of sentences from the text}
$\textbf{var}~sentences$\;
\tcp{set of previously used keywords}
$\textbf{var}~usedKeywords$\;
\ForEach{$s \in sentences$}{
$keywords$ = identifyKeywords(s)\;
$keyword$ = selectFirstNotIn(usedKeywords, keywords)\;
$current$ = getType(keyword)\;

\uIf{isLastSentence($s$)}{
    $child = current.root.children[0]$\;
    $transitionOverviewToFocus(child)$\;
    $transitionFocusToOverview(child)$\;
    $overviewScene(child)$\;
}
\uElseIf{hasChildren($current$)}{
    $transitionFocusToOverview(current)$\;
    $overviewScene(current)$\;
}
\Else{
    \uIf{$previous.parent != current.parent$}{
        $transitionSiblings(current.parent, current)$\;
        $focusScene(current)$\;
    }
    \Else{
        $transitionSiblings(current, current)$\;
        $focusScene(current)$\;
    }
}
$usedKeywords.insert(keyword)$\;
$previous = current$\;
}
\caption{\mbox{~Scene generation (text-to-molecumentary)}}\label{alg:text-to-molecumentary}
\end{algorithm}

\section{Results and Discussion}
We developed a prototypical implementation of the molecumentary synthesis on top of the Marion library~\cite{marion}, which supports biology communication.
Our molecular rendering uses cell\-VIEW's~\cite{cellview} impostor approach, coupled with le\-vels-of-de\-tail for an efficient depiction of large molecular models.

To fetch the descriptive texts for the model elements, we could use any repository with such data.
As noted above, we used Wikipedia's API to get article \emph{extracts}: short descriptions of the keywords. The response time for such a query was $\approx$\,150~ms---well within the limits of a live production.
We found that, in the majority of our tests, three sentences from the extracts are sufficient to describe a structure. The quality of the result, however, highly depends on the quality of the search terms, \ie, the structure identifiers in the annotated model. If the model is not well-annotated or keywords are too general, the results can be unrelated or misleading.

Our framework's component for verbalizing texts is implementation-agnostic. Since Marion is based on Qt, we are able to leverage its text-to-speech functionality. The Qt Speech component~\cite{qt-speech} provides an abstraction layer above text-to-speech interfaces available for several OSes, \eg, \texttt{libspeechd} for Linux or Windows' native library.
As an alternative, we also interface with an online service, Google's Cloud Text-to-Speech API~\cite{google-speech}. It allows us to customize several speech attributes speech and, in our experience, produces more natural sounding speech output.
To support languages other than English, we use a user-defined keyword translation. These translated keywords allow us to retrieve relevant information in the target language.

We tested our approach on three molecular datasets, which we discuss next.
In our supplementary video we show parts from all molecumentaries produced from these datasets, which we recorded in real-time at FullHD resolution.

\begin{figure*}
    \centering
		\setlength{\subfigcapskip}{-4ex}%
    \textcolor{white}{\subfigure[\hspace{.29\linewidth}]{%
	\includegraphics[width=.32\linewidth]{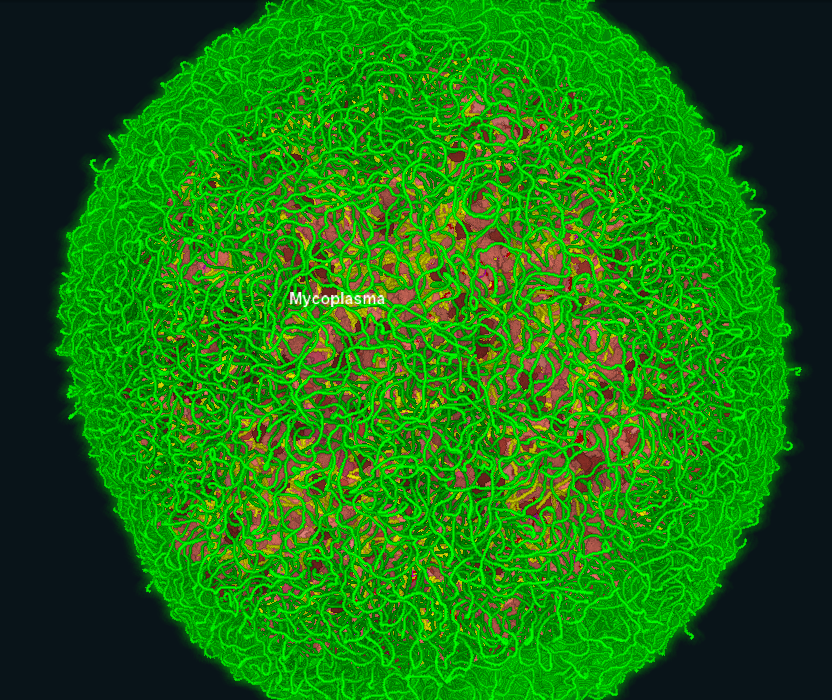}%
	\label{fig:results-myco1}
}}\hfill%
\textcolor{white}{\subfigure[\hspace{.29\linewidth}]{%
	\includegraphics[width=.32\linewidth]{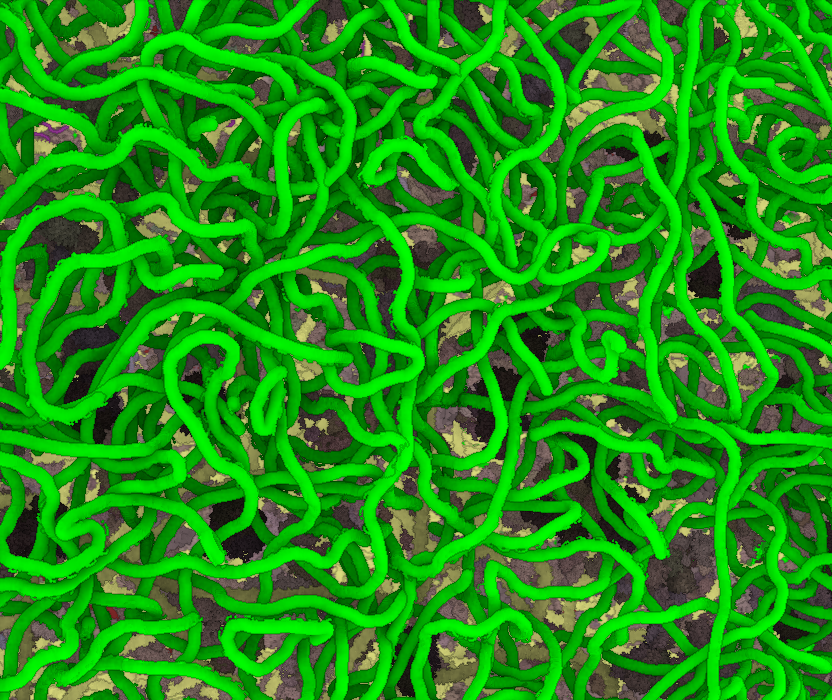}%
	\label{fig:results-myco2}
}}\hfill%
\textcolor{white}{\subfigure[\hspace{.29\linewidth}]{%
	\includegraphics[width=.32\linewidth]{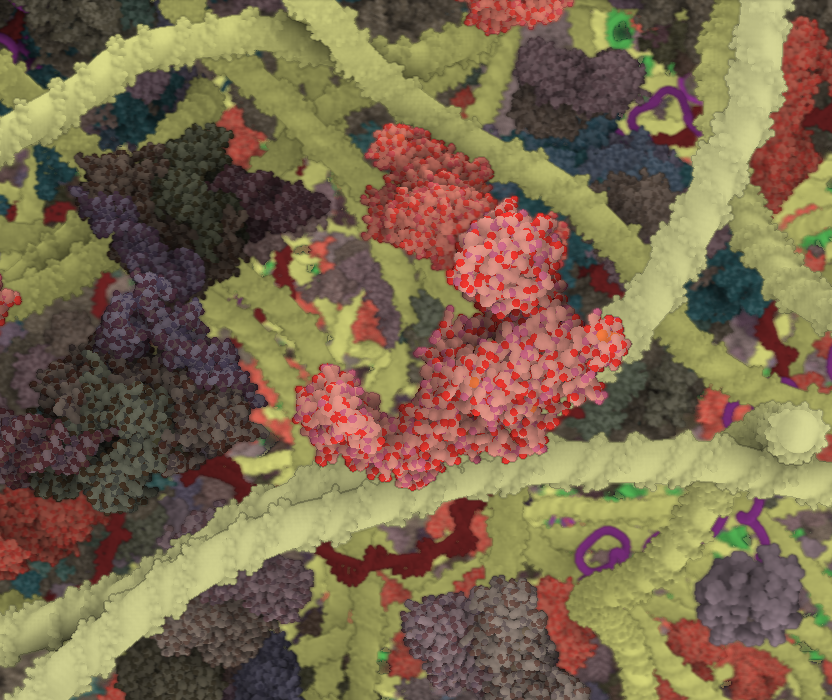}%
	\label{fig:results-myco3}
 }}\vspace{-1ex}
\caption{The Mycoplasma model contains many more fiber instances. Throughout the virtual tour we visit both the strands visible from the outside view (\eg, peptides shown in \subref{fig:results-myco2}) as well as the insides of the bacteria pictured in \subref{fig:results-myco3}.}\vspace{-1ex}
    \label{fig:results-myco}
\end{figure*}

\begin{figure*}
    \centering
		\setlength{\subfigcapskip}{-4ex}%
    \textcolor{white}{\subfigure[\hspace{.29\linewidth}]{%
	\includegraphics[width=.32\linewidth]{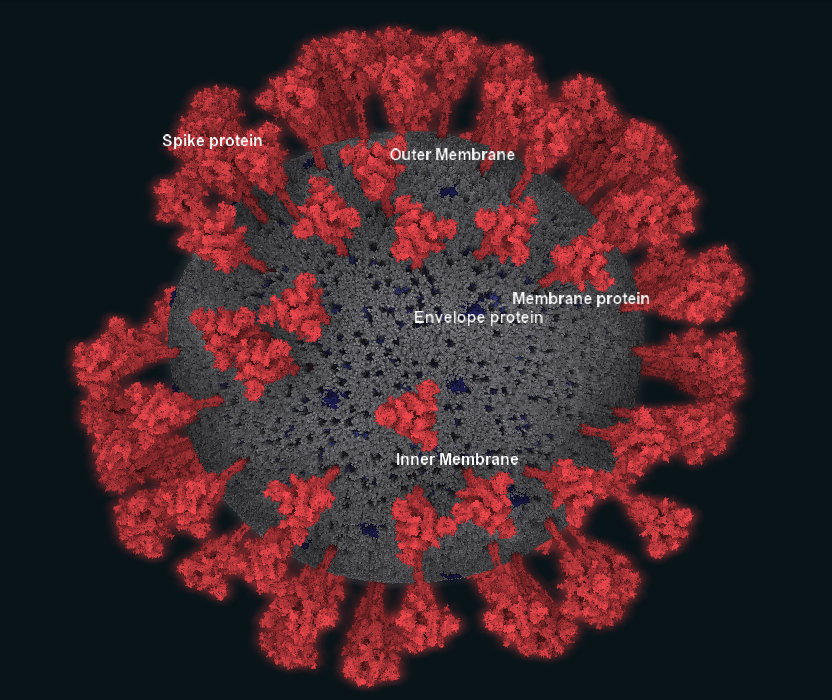}%
	\label{fig:results-covid1}
}}\hfill%
\textcolor{white}{\subfigure[\hspace{.29\linewidth}]{%
	\includegraphics[width=.32\linewidth]{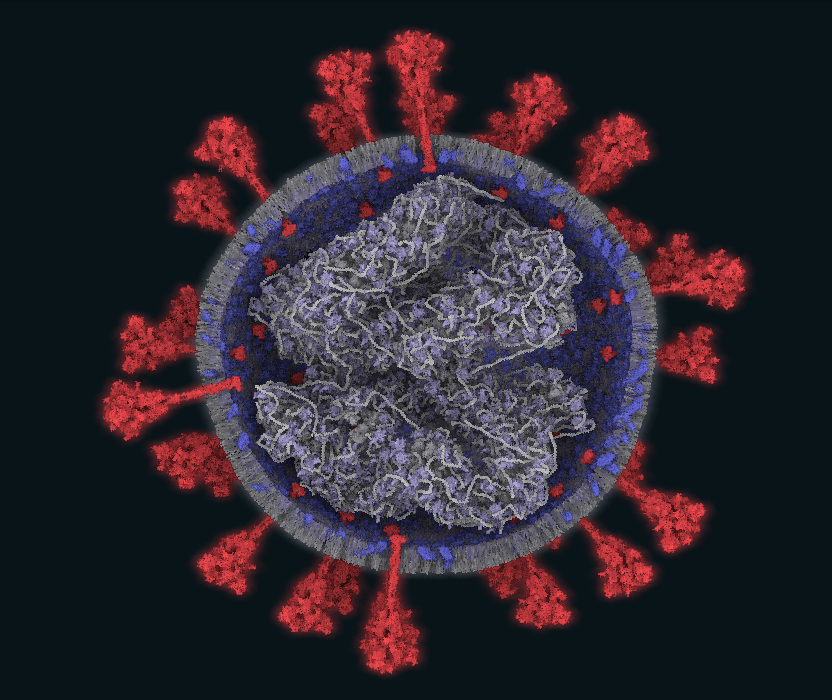}%
	\label{fig:results-covid2}
}}\hfill%
\textcolor{white}{\subfigure[\hspace{.29\linewidth}]{%
	\includegraphics[width=.32\linewidth]{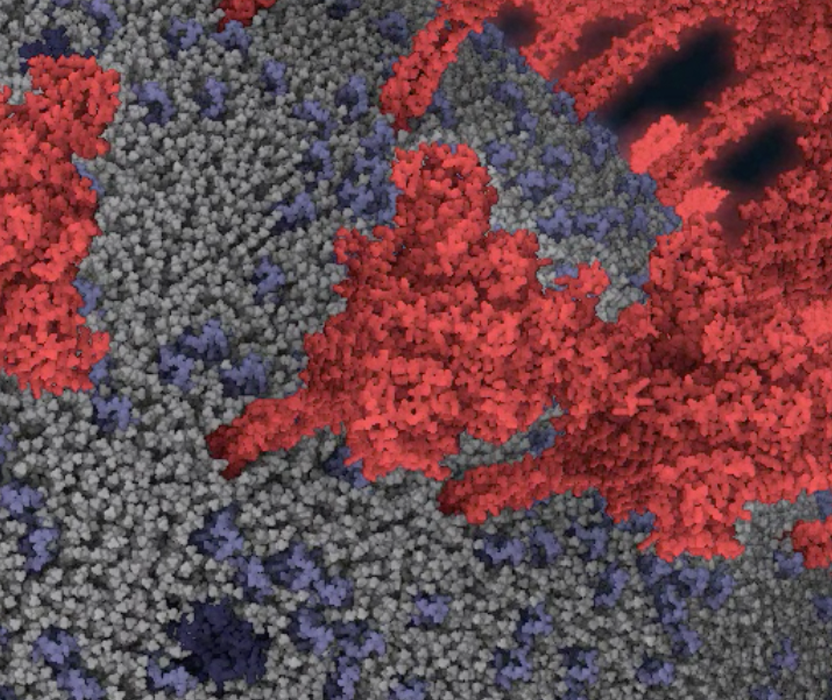}%
	\label{fig:results-covid3}
 }}\vspace{-1ex}
\caption{The SARS-CoV-2 model shows the composition of the virus. We see its inside composition in an overview scene \subref{fig:results-covid2}, and the very important structure---the spike protein---is then shown in detail in a focus scene \subref{fig:results-covid3}.}\vspace{-1ex}
    \label{fig:results-covid}
\end{figure*}

First, we use a \textbf{HIV in blood plasma} dataset (\autoref{fig:results-hiv}) from Scripps Research. It consists of 45 protein types with $\approx$\,18,500 protein instances, 200,000 lipids, and a single RNA strand. Its hierarchy is five levels deep and the model is well annotated with descriptive names. Every molecule type has a human readable name provided by an expert, and almost all of them have a local textual description.

For this model we asked a domain expert to provide a textual description, which we used in the text-to-molecumentary scenario. The resulting molecumentary is 2:42 minutes long and the narration is fluent as it would read by a human narrator. The transition scenes that we inserted between the focus and overview scenes for the sentences do not influence the movie's length.
We also produced a self-gui\-ded movie, which we stopped after 4:33 minutes. In this time, our framework visited 11 story-graph nodes. The synthetic navigational sentences took $\approx$\,26\% of the movie's time.
Using a nVidia Titan V in FullHD resolution, we are able to render both movies at at least 17\,fps (in zoomed scenes), with a median of $\approx$\,27\,fps (overview scenes).

Second, the \textbf{Mycoplasma dataset} (\autoref{fig:results-myco}) has also been provided by Scripps Research. The model comprises $\approx$\,5,400 protein instances of 18 different types and RNA. Because it is a preliminary model it does not yet contain a lipid membrane. Approximately a half of the proteins is well annotated and there is no textual description provided with the model. For that reason we downloaded all needed descriptive texts from Wikipedia and only produced a self-gui\-ded movie. We stopped this movie after visiting 11 story-graph nodes at 4:35 minutes. 24\% of this time was spent on navigational sentences. The performance in FullHD resolution using nVidia Titan V was 14\,fps or more.

Finally, we used the \textbf{SARS-CoV-2} dataset (\autoref{fig:results-covid}) provided by KAUST \cite{nguyen2020modeling}.
It consists of $\approx$\,3,200 protein instances of six different protein types, $\approx$\,180,000 lipids, and an RNA strand. The model is annotated with human readable labels, but no predefined textual description is available.
We thus also retrieved all textual descriptions from Wikipedia and only produced a self-guided movie.
It is 3:20 minutes long, visiting 10 story-graph nodes. The performance at FullHD resolution with the nVidia Titan V is 20\,fps or more. The navigational sentences took $\approx$\,31\% of the movie. With this dataset we discovered an aspect that needs improvement: Because the leaves in the hierarchy are individual RNA bases that consist of only a few atoms, the camera traversal while zooming in goes too much into the depth. The resulting view is not very attractive, and we need to address it in the future.

To reflect on our work and validate its utility in biology communication, we showed these results to two domain experts, with  45 resp.\ 33 years of professional experience.
They appreciated its potential in being able to showcase complex molecular models
and confirmed that the many parts of these models are difficult to understand with conventional interaction methods.
Both domain experts also liked the coordination of the generated speech with the visual content, commenting that the transitions are easy to follow.
They also noted that our method would make a valuable tool for semi-automated content creation, provided that we add more user interaction in the creation pipeline.

The domain experts also pointed out some limitations.
In particular, the final zoomed-in view does not always end up showing the molecules from a characteristic view. To solve this issue, canonical views of each structural type could be computed and used in the determination of the final camera position.
Furthermore, to simplify the design of our method, we only focus on a single component at a time. One domain expert mentioned that it would be good to be able to explain two (or more) components at a time and include a commentary of their interaction.
Finally, we considered only static models so far. Models from molecular dynamics simulations would present additional challenges.

\section{Conclusion and Future Work}
In the domain of molecular and biological visualization there is a movement towards combining data from various sources and contextualizing them in a single environment~\cite{autin:2020:mesoscope}. The goal is to develop a pipeline that automates the whole process from data acquisition and modeling, to visualization and rendering.
Our approach contributes to this effort and we consider our framework to be the initial step toward automatic interactive storytelling in the context of science communication. We can automatically integrate semantic information---fetched from online sources or provided by experts---about the composition of a molecular model. Our work is made possible by the advances of real-time visualization. Real-time graphics, as opposed to offline rendering approaches, is being rapidly utilized in moviemaking and we believe that adopting a similar trend in visualization can fundamentally change the field of scientific outreach. Yet the field of molecular visualization still lacks sufficient standardization that would allow us to create a fully automated pipeline from observation to science communication.

Nonetheless, with our work we still contribute to the latter field of scientific outreach.
While we cannot and do not intend to replace domain experts who explain specific concepts (\ie, the science communicator), with our current technology we can take advantage of the same sources that experts use, extract the key information, and deploy it on-demand to an audience at any time. As such we are able to provide visually supported scientific narratives where it was not possible to use them before, in a similar way that illustrative visualization allows us to use illustration-like visuals where we cannot afford human illustrators.

Many directions are possible for future work. We are interested in exploring the idea of the \emph{interaction spectrum}.
One end of the spectrum corresponds to fully interactive control. The other end corresponds to passive viewing without interaction. In-between, various levels of constrained navigation and guidance are worth exploring.
The incorporation of artifical speech technology also suggests to exploit the opposite direction: parsing a human speech and letting the spectator's words influence the interactive experience or, in our case, the narrative of the scientific documentary.

\ifCLASSOPTIONcaptionsoff
  \newpage
\fi



\bibliographystyle{IEEEtranS}
\bibliography{template}

\begin{thebibliography}{10}
\providecommand{\url}[1]{#1}
\csname url@samestyle\endcsname
\providecommand{\newblock}{\relax}
\providecommand{\bibinfo}[2]{#2}
\providecommand{\BIBentrySTDinterwordspacing}{\spaceskip=0pt\relax}
\providecommand{\BIBentryALTinterwordstretchfactor}{4}
\providecommand{\BIBentryALTinterwordspacing}{\spaceskip=\fontdimen2\font plus
\BIBentryALTinterwordstretchfactor\fontdimen3\font minus
  \fontdimen4\font\relax}
\providecommand{\BIBforeignlanguage}[2]{{%
\expandafter\ifx\csname l@#1\endcsname\relax
\typeout{** WARNING: IEEEtranS.bst: No hyphenation pattern has been}%
\typeout{** loaded for the language `#1'. Using the pattern for}%
\typeout{** the default language instead.}%
\else
\language=\csname l@#1\endcsname
\fi
#2}}
\providecommand{\BIBdecl}{\relax}
\BIBdecl

\bibitem{ahmed-2005:auto-camera-graph-nav}
\href{https://dl.acm.org/doi/10.5555/1082315.1082320}{A.~Ahmed and P.~Eades},
  \href{https://dl.acm.org/doi/10.5555/1082315.1082320}{``Automatic camera path
  generation for graph navigation in {3D}},''
  \href{https://dl.acm.org/doi/10.5555/1082315.1082320}{in \emph{Proc.\
  APVis}},
  \href{https://dl.acm.org/doi/10.5555/1082315.1082320}{vol.~45}.\hskip 1em
  plus 0.5em minus 0.4em\relax
  \href{https://dl.acm.org/doi/10.5555/1082315.1082320}{Australia: Australian
  Computer Society, Inc., 2005},
  \href{https://dl.acm.org/doi/10.5555/1082315.1082320}{pp. 27--32}.

\bibitem{akiba:2010:aniviz}
\href{https://doi.org/10.1109/MCG.2009.107}{H.~{Akiba}, C.~{Wang}, and K.-L.
  {Ma}}, \href{https://doi.org/10.1109/MCG.2009.107}{``Aniviz: A template-based
  animation tool for volume visualization},''
  \href{https://doi.org/10.1109/MCG.2009.107}{\emph{IEEE Computer Graphics and
  Applications}}, \href{https://doi.org/10.1109/MCG.2009.107}{vol.~30},
  \href{https://doi.org/10.1109/MCG.2009.107}{no.~5},
  \href{https://doi.org/10.1109/MCG.2009.107}{pp. 61--71},
  \href{https://doi.org/10.1109/MCG.2009.107}{2010}.
  \href{https://doi.org/10.1109/MCG.2009.107}
{doi: {{%
10\hspace{.1pt}\discretionary{.}{%
}{.}\hspace{.4pt}1109\discretionary{/}{%
}{/}MCG\hspace{.1pt}\discretionary{.}{%
}{.}\hspace{.4pt}2009\hspace{.1pt}\discretionary{.}{%
}{.}\hspace{.4pt}107}}}


\bibitem{Amini2015}
\href{https://doi.org/10.1145/2702123.2702431}{F.~Amini, N.~Henry~Riche,
  B.~Lee, C.~Hurter, and P.~Irani},
  \href{https://doi.org/10.1145/2702123.2702431}{``Understanding data videos:
  Looking at narrative visualization through the cinematography lens},''
  \href{https://doi.org/10.1145/2702123.2702431}{in \emph{Proc.\ CHI}}.\hskip
  1em plus 0.5em minus 0.4em\relax
  \href{https://doi.org/10.1145/2702123.2702431}{New York: ACM, 2015},
  \href{https://doi.org/10.1145/2702123.2702431}{pp. 1459--1468}.
  \href{https://doi.org/10.1145/2702123.2702431}
{doi: {{%
10\hspace{.1pt}\discretionary{.}{%
}{.}\hspace{.4pt}1145\discretionary{/}{%
}{/}2702123\hspace{.1pt}\discretionary{.}{%
}{.}\hspace{.4pt}2702431}}}


\bibitem{autin:2020:mesoscope}
\href{https://doi.org/10.2312/molva.20201098}{L.~Autin, M.~Maritan, B.~A.
  Barbaro, A.~Gardner, A.~J. Olson, M.~Sanner, and D.~S. Goodsell},
  \href{https://doi.org/10.2312/molva.20201098}{``Mesoscope: A web-based tool
  for mesoscale data integration and curation},''
  \href{https://doi.org/10.2312/molva.20201098}{in \emph{Proc.\ MolVA}}.\hskip
  1em plus 0.5em minus 0.4em\relax
  \href{https://doi.org/10.2312/molva.20201098}{Goslar, Germany: Eurographics
  Assoc., 2020}, \href{https://doi.org/10.2312/molva.20201098}{pp. 23--31}.
  \href{https://doi.org/10.2312/molva.20201098}
{doi: {{%
10\hspace{.1pt}\discretionary{.}{%
}{.}\hspace{.4pt}2312\discretionary{/}{%
}{/}molva\hspace{.1pt}\discretionary{.}{%
}{.}\hspace{.4pt}20201098}}}


\bibitem{bock:2020:openspace-new}
\href{https://doi.org/10.1109/TVCG.2019.2934259}{A.~{Bock}, E.~{Axelsson},
  J.~{Costa}, G.~{Payne}, M.~{Acinapura}, V.~{Trakinski}, C.~{Emmart},
  C.~{Silva}, C.~{Hansen}, and A.~{Ynnerman}},
  \href{https://doi.org/10.1109/TVCG.2019.2934259}{``{OpenSpace}: A system for
  astrographics},'' \href{https://doi.org/10.1109/TVCG.2019.2934259}{\emph{IEEE
  Transactions on Visualization and Computer Graphics}},
  \href{https://doi.org/10.1109/TVCG.2019.2934259}{vol.~26},
  \href{https://doi.org/10.1109/TVCG.2019.2934259}{no.~1},
  \href{https://doi.org/10.1109/TVCG.2019.2934259}{pp. 633--642},
  \href{https://doi.org/10.1109/TVCG.2019.2934259}{Jan 2020}.
  \href{https://doi.org/10.1109/TVCG.2019.2934259}
{doi: {{%
10\hspace{.1pt}\discretionary{.}{%
}{.}\hspace{.4pt}1109\discretionary{/}{%
}{/}TVCG\hspace{.1pt}\discretionary{.}{%
}{.}\hspace{.4pt}2019\hspace{.1pt}\discretionary{.}{%
}{.}\hspace{.4pt}2934259}}}


\bibitem{Burtnyk:2002:StyleCam}
\href{https://doi.org/10.1145/571985.572000}{N.~Burtnyk, A.~Khan,
  G.~Fitzmaurice, R.~Balakrishnan, and G.~Kurtenbach},
  \href{https://doi.org/10.1145/571985.572000}{``Style{C}am: Interactive
  stylized {3D} navigation using integrated spatial \& temporal controls},''
  \href{https://doi.org/10.1145/571985.572000}{in \emph{Proc.\ UIST}}.\hskip
  1em plus 0.5em minus 0.4em\relax
  \href{https://doi.org/10.1145/571985.572000}{New York: ACM, 2002},
  \href{https://doi.org/10.1145/571985.572000}{pp. 101--110}.
  \href{https://doi.org/10.1145/571985.572000}
{doi: {{%
10\hspace{.1pt}\discretionary{.}{%
}{.}\hspace{.4pt}1145\discretionary{/}{%
}{/}571985\hspace{.1pt}\discretionary{.}{%
}{.}\hspace{.4pt}572000}}}


\bibitem{Burtnyk:2006:ShowMotion}
\href{https://doi.org/10.1145/1111411.1111442}{N.~Burtnyk, A.~Khan,
  G.~Fitzmaurice, and G.~Kurtenbach},
  \href{https://doi.org/10.1145/1111411.1111442}{``Show{M}otion: Camera motion
  based {3D} design review},''
  \href{https://doi.org/10.1145/1111411.1111442}{in \emph{Proc.\ I3D}}.\hskip
  1em plus 0.5em minus 0.4em\relax
  \href{https://doi.org/10.1145/1111411.1111442}{New York: ACM, 2006},
  \href{https://doi.org/10.1145/1111411.1111442}{pp. 167--174}.
  \href{https://doi.org/10.1145/1111411.1111442}
{doi: {{%
10\hspace{.1pt}\discretionary{.}{%
}{.}\hspace{.4pt}1145\discretionary{/}{%
}{/}1111411\hspace{.1pt}\discretionary{.}{%
}{.}\hspace{.4pt}1111442}}}


\bibitem{christie:2005:survey}
\href{https://doi.org/10.1007/11536482_4}{M.~Christie, R.~Machap, J.-M.
  Normand, P.~Olivier, and J.~Pickering},
  \href{https://doi.org/10.1007/11536482_4}{``Virtual camera planning: A
  survey},'' \href{https://doi.org/10.1007/11536482_4}{in \emph{Proc.\ Smart
  Graphics}}.\hskip 1em plus 0.5em minus 0.4em\relax
  \href{https://doi.org/10.1007/11536482_4}{Berlin, Heidelberg: Springer,
  2005}, \href{https://doi.org/10.1007/11536482_4}{pp. 40--52}.
  \href{https://doi.org/10.1007/11536482_4}
{doi: {{%
10\hspace{.1pt}\discretionary{.}{%
}{.}\hspace{.4pt}1007\discretionary{/}{%
}{/}11536482\_4}}}


\bibitem{christie:2008:overview}
\href{https://doi.org/10.1111/j.1467-8659.2008.01181.x}{M.~Christie,
  P.~Olivier, and J.-M. Normand},
  \href{https://doi.org/10.1111/j.1467-8659.2008.01181.x}{``Camera control in
  computer graphics},''
  \href{https://doi.org/10.1111/j.1467-8659.2008.01181.x}{\emph{Computer
  Graphics Forum}},
  \href{https://doi.org/10.1111/j.1467-8659.2008.01181.x}{vol.~27},
  \href{https://doi.org/10.1111/j.1467-8659.2008.01181.x}{no.~8},
  \href{https://doi.org/10.1111/j.1467-8659.2008.01181.x}{pp. 2197--2218},
  \href{https://doi.org/10.1111/j.1467-8659.2008.01181.x}{2008}.
  \href{https://doi.org/10.1111/j.1467-8659.2008.01181.x}
{doi: {{%
10\hspace{.1pt}\discretionary{.}{%
}{.}\hspace{.4pt}1111\discretionary{/}{%
}{/}j\hspace{.1pt}\discretionary{.}{%
}{.}\hspace{.4pt}1467\discretionary{%
}{-}{-}8659\hspace{.1pt}\discretionary{.}{%
}{.}\hspace{.4pt}2008\hspace{.1pt}\discretionary{.}{%
}{.}\hspace{.4pt}01181\hspace{.1pt}\discretionary{.}{%
}{.}\hspace{.4pt}x}}}


\bibitem{qt-speech}
Q.~Company, ``Qt speech,'' Web site,
  \url{https://doc.qt.io/qt-5/qtspeech-index.html}, accessed July 2020.

\bibitem{Daly:2014:microscopy-movies}
\href{http://eprints.hud.ac.uk/id/eprint/23181}{C.~Daly, L.~Clunie, and M.~Ma},
  \href{http://eprints.hud.ac.uk/id/eprint/23181}{``From microscope to movies:
  {3D} animations for teaching physiology},''
  \href{http://eprints.hud.ac.uk/id/eprint/23181}{\emph{Microscopy and
  Analysis}}, \href{http://eprints.hud.ac.uk/id/eprint/23181}{vol.~28},
  \href{http://eprints.hud.ac.uk/id/eprint/23181}{no.~6},
  \href{http://eprints.hud.ac.uk/id/eprint/23181}{pp. 7--10},
  \href{http://eprints.hud.ac.uk/id/eprint/23181}{Sep./Oct. 2014}.

\bibitem{galvane:2018:directing}
\href{https://doi.org/10.1145/3181975}{Q.~Galvane, C.~Lino, M.~Christie,
  J.~Fleureau, F.~Servant, F.-L. Tariolle, and P.~Guillotel},
  \href{https://doi.org/10.1145/3181975}{``Directing cinematographic drones},''
  \href{https://doi.org/10.1145/3181975}{\emph{{ACM Transactions on
  Graphics}}}, \href{https://doi.org/10.1145/3181975}{vol.~37},
  \href{https://doi.org/10.1145/3181975}{no.~3},
  \href{https://doi.org/10.1145/3181975}{pp. 34:1--34:18},
  \href{https://doi.org/10.1145/3181975}{Aug. 2018}.
  \href{https://doi.org/10.1145/3181975}
{doi: {{%
10\hspace{.1pt}\discretionary{.}{%
}{.}\hspace{.4pt}1145\discretionary{/}{%
}{/}3181975}}}


\bibitem{Gershon2001}
\href{https://doi.org/10.1145/381641.381653}{N.~Gershon and W.~Page},
  \href{https://doi.org/10.1145/381641.381653}{``What storytelling can do for
  information visualization},''
  \href{https://doi.org/10.1145/381641.381653}{\emph{Communications of the
  ACM}}, \href{https://doi.org/10.1145/381641.381653}{vol.~44},
  \href{https://doi.org/10.1145/381641.381653}{no.~8},
  \href{https://doi.org/10.1145/381641.381653}{pp. 31--37},
  \href{https://doi.org/10.1145/381641.381653}{Aug. 2001}.
  \href{https://doi.org/10.1145/381641.381653}
{doi: {{%
10\hspace{.1pt}\discretionary{.}{%
}{.}\hspace{.4pt}1145\discretionary{/}{%
}{/}381641\hspace{.1pt}\discretionary{.}{%
}{.}\hspace{.4pt}381653}}}


\bibitem{google-speech}
Google, ``Cloud text-to-speech {API},''
  \url{https://cloud.google.com/text-to-speech/docs/reference/rest/}, accessed
  July 2020.

\bibitem{gratzl2016clue}
\href{https://doi.org/10.1111/cgf.12925}{S.~Gratzl, A.~Lex, N.~Gehlenborg,
  N.~Cosgrove, and M.~Streit}, \href{https://doi.org/10.1111/cgf.12925}{``From
  visual exploration to storytelling and back again},''
  \href{https://doi.org/10.1111/cgf.12925}{\emph{Computer Graphics Forum}},
  \href{https://doi.org/10.1111/cgf.12925}{vol.~35},
  \href{https://doi.org/10.1111/cgf.12925}{no.~3},
  \href{https://doi.org/10.1111/cgf.12925}{pp. 491--500},
  \href{https://doi.org/10.1111/cgf.12925}{Jun. 2016}.
  \href{https://doi.org/10.1111/cgf.12925}
{doi: {{%
10\hspace{.1pt}\discretionary{.}{%
}{.}\hspace{.4pt}1111\discretionary{/}{%
}{/}cgf\hspace{.1pt}\discretionary{.}{%
}{.}\hspace{.4pt}12925}}}


\bibitem{Hsu2011}
\href{https://doi.org/10.1145/2070781.2024165}{W.-H. Hsu, K.-L. Ma, and
  C.~Correa}, \href{https://doi.org/10.1145/2070781.2024165}{``A rendering
  framework for multiscale views of {3D} models},''
  \href{https://doi.org/10.1145/2070781.2024165}{\emph{ACM Transactions on
  Graphics}}, \href{https://doi.org/10.1145/2070781.2024165}{vol.~30},
  \href{https://doi.org/10.1145/2070781.2024165}{no.~6},
  \href{https://doi.org/10.1145/2070781.2024165}{pp. 1--10},
  \href{https://doi.org/10.1145/2070781.2024165}{Dec. 2011}.
  \href{https://doi.org/10.1145/2070781.2024165}
{doi: {{%
10\hspace{.1pt}\discretionary{.}{%
}{.}\hspace{.4pt}1145\discretionary{/}{%
}{/}2070781\hspace{.1pt}\discretionary{.}{%
}{.}\hspace{.4pt}2024165}}}


\bibitem{Hullman2011}
\href{https://doi.org/10.1109/TVCG.2011.255}{J.~Hullman and N.~Diakopoulos},
  \href{https://doi.org/10.1109/TVCG.2011.255}{``Visualization rhetoric:
  Framing effects in narrative visualization},''
  \href{https://doi.org/10.1109/TVCG.2011.255}{\emph{IEEE Transactions on
  Visualization and Computer Graphics}},
  \href{https://doi.org/10.1109/TVCG.2011.255}{vol.~17},
  \href{https://doi.org/10.1109/TVCG.2011.255}{no.~12},
  \href{https://doi.org/10.1109/TVCG.2011.255}{pp. 2231--2240},
  \href{https://doi.org/10.1109/TVCG.2011.255}{Dec. 2011}.
  \href{https://doi.org/10.1109/TVCG.2011.255}
{doi: {{%
10\hspace{.1pt}\discretionary{.}{%
}{.}\hspace{.4pt}1109\discretionary{/}{%
}{/}TVCG\hspace{.1pt}\discretionary{.}{%
}{.}\hspace{.4pt}2011\hspace{.1pt}\discretionary{.}{%
}{.}\hspace{.4pt}255}}}


\bibitem{cellpack}
\href{https://doi.org/10.1038/nmeth.3204}{G.~T. Johnson, L.~Autin, M.~Al-Alusi,
  D.~S. Goodsell, M.~F. Sanner, and A.~J. Olson},
  \href{https://doi.org/10.1038/nmeth.3204}{``{cell{P}{A}{C}{K}: A virtual
  mesoscope to model and visualize structural systems biology}},''
  \href{https://doi.org/10.1038/nmeth.3204}{\emph{Nature Methods}},
  \href{https://doi.org/10.1038/nmeth.3204}{vol.~12},
  \href{https://doi.org/10.1038/nmeth.3204}{no.~1},
  \href{https://doi.org/10.1038/nmeth.3204}{pp. 85--91},
  \href{https://doi.org/10.1038/nmeth.3204}{Dec. 2015}.
  \href{https://doi.org/10.1038/nmeth.3204}
{doi: {{%
10\hspace{.1pt}\discretionary{.}{%
}{.}\hspace{.4pt}1038\discretionary{/}{%
}{/}nmeth\hspace{.1pt}\discretionary{.}{%
}{.}\hspace{.4pt}3204}}}


\bibitem{cellPACK_HIV_Johnson2014}
\href{https://doi.org/10.1039/c4fd00017j}{G.~T. Johnson, D.~S. Goodsell,
  L.~Autin, S.~Forli, M.~F. Sanner, and A.~J. Olson},
  \href{https://doi.org/10.1039/c4fd00017j}{``\BIBforeignlanguage{eng}{{3D
  molecular models of whole HIV-1 virions generated with cellPACK}}},''
  \href{https://doi.org/10.1039/c4fd00017j}{\emph{\BIBforeignlanguage{eng}{Faraday
  Discussions}}}, \href{https://doi.org/10.1039/c4fd00017j}{vol. 169},
  \href{https://doi.org/10.1039/c4fd00017j}{pp. 23--44},
  \href{https://doi.org/10.1039/c4fd00017j}{Sep. 2014}.
  \href{https://doi.org/10.1039/c4fd00017j}
{doi: {{%
10\hspace{.1pt}\discretionary{.}{%
}{.}\hspace{.4pt}1039\discretionary{/}{%
}{/}c4fd00017j}}}


\bibitem{Karpe2018}
\href{https://doi.org/10.22214/ijraset.2018.3054}{R.~Karpe},
  \href{https://doi.org/10.22214/ijraset.2018.3054}{``A survey :{O}n text to
  speech synthesis},''
  \href{https://doi.org/10.22214/ijraset.2018.3054}{\emph{International Journal
  for Research in Applied Science and Engineering Technology}},
  \href{https://doi.org/10.22214/ijraset.2018.3054}{vol.~6},
  \href{https://doi.org/10.22214/ijraset.2018.3054}{no.~03},
  \href{https://doi.org/10.22214/ijraset.2018.3054}{pp. 351--355},
  \href{https://doi.org/10.22214/ijraset.2018.3054}{Mar. 2018}.
  \href{https://doi.org/10.22214/ijraset.2018.3054}
{doi: {{%
10\hspace{.1pt}\discretionary{.}{%
}{.}\hspace{.4pt}22214\discretionary{/}{%
}{/}ijraset\hspace{.1pt}\discretionary{.}{%
}{.}\hspace{.4pt}2018\hspace{.1pt}\discretionary{.}{%
}{.}\hspace{.4pt}3054}}}


\bibitem{Knoebelreiter:2014:architectural-flythrough}
\href{https://doi.org/10.5220/0004670303350341}{P.~{Kn{\"o}belreiter},
  R.~{Berndt}, T.~{Ullrich}, and D.~W. {Fellner}},
  \href{https://doi.org/10.5220/0004670303350341}{``Automatic fly-through
  camera animations for {3D} architectural repositories},''
  \href{https://doi.org/10.5220/0004670303350341}{in \emph{Proc.\
  GRAPP}}.\hskip 1em plus 0.5em minus 0.4em\relax
  \href{https://doi.org/10.5220/0004670303350341}{IEEE, 2014},
  \href{https://doi.org/10.5220/0004670303350341}{pp. 335--341}.
  \href{https://doi.org/10.5220/0004670303350341}
{doi: {{%
10\hspace{.1pt}\discretionary{.}{%
}{.}\hspace{.4pt}5220\discretionary{/}{%
}{/}0004670303350341}}}


\bibitem{kosara2013}
\href{https://doi.org/10.1109/MC.2013.36}{R.~{Kosara} and J.~{Mackinlay}},
  \href{https://doi.org/10.1109/MC.2013.36}{``Storytelling: The next step for
  visualization},'' \href{https://doi.org/10.1109/MC.2013.36}{\emph{IEEE
  Computer}}, \href{https://doi.org/10.1109/MC.2013.36}{vol.~46},
  \href{https://doi.org/10.1109/MC.2013.36}{no.~5},
  \href{https://doi.org/10.1109/MC.2013.36}{pp. 44--50},
  \href{https://doi.org/10.1109/MC.2013.36}{May 2013}.
  \href{https://doi.org/10.1109/MC.2013.36}
{doi: {{%
10\hspace{.1pt}\discretionary{.}{%
}{.}\hspace{.4pt}1109\discretionary{/}{%
}{/}MC\hspace{.1pt}\discretionary{.}{%
}{.}\hspace{.4pt}2013\hspace{.1pt}\discretionary{.}{%
}{.}\hspace{.4pt}36}}}


\bibitem{kouril:2020:hyperlabels}
\href{https://doi.org/10.1109/TVCG.2020.2975583}{D.~{Kou\v{r}il},
  T.~{Isenberg}, B.~{Kozl\'{i}kov\'{a}}, M.~{Meyer}, E.~{Gr{\"o}ller}, and
  I.~{Viola}},
  \href{https://doi.org/10.1109/TVCG.2020.2975583}{``{HyperLabels}: Browsing of
  dense and hierarchical molecular {3D} models},''
  \href{https://doi.org/10.1109/TVCG.2020.2975583}{\emph{IEEE Transactions on
  Visualization and Computer Graphics}},
  \href{https://doi.org/10.1109/TVCG.2020.2975583}{2020},
  \href{https://doi.org/10.1109/TVCG.2020.2975583}{to appear}.
  \href{https://doi.org/10.1109/TVCG.2020.2975583}
{doi: {{%
10\hspace{.1pt}\discretionary{.}{%
}{.}\hspace{.4pt}1109\discretionary{/}{%
}{/}TVCG\hspace{.1pt}\discretionary{.}{%
}{.}\hspace{.4pt}2020\hspace{.1pt}\discretionary{.}{%
}{.}\hspace{.4pt}2975583}}}


\bibitem{kwon2014}
\href{https://www.microsoft.com/en-us/research/publication/visjockey-enriching-data-stories-orchestrated-interactive-visualization/}{B.~C.
  Kwon, F.~Stoffel, D.~J{\"a}ckle, B.~Lee, and D.~Keim},
  \href{https://www.microsoft.com/en-us/research/publication/visjockey-enriching-data-stories-orchestrated-interactive-visualization/}{``Vis{J}ockey:
  Enriching data stories through orchestrated interactive visualization},''
  \href{https://www.microsoft.com/en-us/research/publication/visjockey-enriching-data-stories-orchestrated-interactive-visualization/}{in
  \emph{Proc.\ Computation+Journalism Symp.}}\hskip 1em plus 0.5em minus
  0.4em\relax
  \href{https://www.microsoft.com/en-us/research/publication/visjockey-enriching-data-stories-orchestrated-interactive-visualization/}{New
  York: Brown Institute for Media Innovation, 2014}.

\bibitem{lee2015}
\href{https://doi.org/10.1109/MCG.2015.99}{B.~{Lee}, N.~H. {Riche},
  P.~{Isenberg}, and S.~{Carpendale}},
  \href{https://doi.org/10.1109/MCG.2015.99}{``More than telling a story:
  Transforming data into visually shared stories},''
  \href{https://doi.org/10.1109/MCG.2015.99}{\emph{IEEE Computer Graphics and
  Applications}}, \href{https://doi.org/10.1109/MCG.2015.99}{vol.~35},
  \href{https://doi.org/10.1109/MCG.2015.99}{no.~5},
  \href{https://doi.org/10.1109/MCG.2015.99}{pp. 84--90},
  \href{https://doi.org/10.1109/MCG.2015.99}{Sep./Oct. 2015}.
  \href{https://doi.org/10.1109/MCG.2015.99}
{doi: {{%
10\hspace{.1pt}\discretionary{.}{%
}{.}\hspace{.4pt}1109\discretionary{/}{%
}{/}MCG\hspace{.1pt}\discretionary{.}{%
}{.}\hspace{.4pt}2015\hspace{.1pt}\discretionary{.}{%
}{.}\hspace{.4pt}99}}}


\bibitem{liao2014}
\href{https://doi.org/10.1007/978-3-319-11650-1_1}{I.~Liao, W.-H. Hsu, and
  K.-L. Ma}, \href{https://doi.org/10.1007/978-3-319-11650-1_1}{``Storytelling
  via navigation: A novel approach to animation for scientific
  visualization},'' \href{https://doi.org/10.1007/978-3-319-11650-1_1}{in
  \emph{Proc.\ Smart Graphics}}.\hskip 1em plus 0.5em minus 0.4em\relax
  \href{https://doi.org/10.1007/978-3-319-11650-1_1}{Cham, Switzerland:
  Springer, 2014}, \href{https://doi.org/10.1007/978-3-319-11650-1_1}{pp.
  1--14}. \href{https://doi.org/10.1007/978-3-319-11650-1_1}
{doi: {{%
10\hspace{.1pt}\discretionary{.}{%
}{.}\hspace{.4pt}1007\discretionary{/}{%
}{/}978\discretionary{%
}{-}{-}3\discretionary{%
}{-}{-}319\discretionary{%
}{-}{-}11650\discretionary{%
}{-}{-}1\_1}}}


\bibitem{lidal2012}
\href{https://doi.org/10.2312/SBM/SBM12/011-020}{E.~M. Lidal, H.~Hauser, and
  I.~Viola}, \href{https://doi.org/10.2312/SBM/SBM12/011-020}{``Geological
  storytelling -- {G}raphically exploring and communicating geological
  sketches},'' \href{https://doi.org/10.2312/SBM/SBM12/011-020}{in \emph{Proc.\
  SBIM}}.\hskip 1em plus 0.5em minus 0.4em\relax
  \href{https://doi.org/10.2312/SBM/SBM12/011-020}{Goslar, Germany:
  Eurographics Assoc., 2012},
  \href{https://doi.org/10.2312/SBM/SBM12/011-020}{pp. 11--20}.
  \href{https://doi.org/10.2312/SBM/SBM12/011-020}
{doi: {{%
10\hspace{.1pt}\discretionary{.}{%
}{.}\hspace{.4pt}2312\discretionary{/}{%
}{/}SBM\discretionary{/}{%
}{/}SBM12\discretionary{/}{%
}{/}011\discretionary{%
}{-}{-}020}}}


\bibitem{lino:2010:cinematography}
\href{https://doi.org/10.2312/SCA/SCA10/139-148}{C.~Lino, M.~Christie,
  F.~Lamarche, G.~Schofield, and P.~Olivier},
  \href{https://doi.org/10.2312/SCA/SCA10/139-148}{``A real-time cinematography
  system for interactive {3D} environments},''
  \href{https://doi.org/10.2312/SCA/SCA10/139-148}{in \emph{Proc.\ SCA}}.\hskip
  1em plus 0.5em minus 0.4em\relax
  \href{https://doi.org/10.2312/SCA/SCA10/139-148}{Goslar, Germany:
  Eurographics Assoc., 2010},
  \href{https://doi.org/10.2312/SCA/SCA10/139-148}{pp. 139--148}.
  \href{https://doi.org/10.2312/SCA/SCA10/139-148}
{doi: {{%
10\hspace{.1pt}\discretionary{.}{%
}{.}\hspace{.4pt}2312\discretionary{/}{%
}{/}SCA\discretionary{/}{%
}{/}SCA10\discretionary{/}{%
}{/}139\discretionary{%
}{-}{-}148}}}


\bibitem{liu:2020:autocaption}
\href{https://doi.org/10.1109/PacificVis48177.2020.1043}{C.~{Liu}, L.~{Xie},
  Y.~{Han}, D.~{Wei}, and X.~{Yuan}},
  \href{https://doi.org/10.1109/PacificVis48177.2020.1043}{``{AutoCaption}: An
  approach to generate natural language description from visualization
  automatically},'' \href{https://doi.org/10.1109/PacificVis48177.2020.1043}{in
  \emph{Proc.\ PacificVis}}.\hskip 1em plus 0.5em minus 0.4em\relax
  \href{https://doi.org/10.1109/PacificVis48177.2020.1043}{Los Alamitos: IEEE
  Computer Society, 2020},
  \href{https://doi.org/10.1109/PacificVis48177.2020.1043}{pp. 191--195}.
  \href{https://doi.org/10.1109/PacificVis48177.2020.1043}
{doi: {{%
10\hspace{.1pt}\discretionary{.}{%
}{.}\hspace{.4pt}1109\discretionary{/}{%
}{/}PacificVis48177\hspace{.1pt}\discretionary{.}{%
}{.}\hspace{.4pt}2020\hspace{.1pt}\discretionary{.}{%
}{.}\hspace{.4pt}1043}}}


\bibitem{Ma2012}
\href{https://doi.org/10.1109/MCG.2012.24}{K.-L. Ma, I.~Liao, J.~Frazier,
  H.~Hauser, and H.~N. Kostis},
  \href{https://doi.org/10.1109/MCG.2012.24}{``Scientific storytelling using
  visualization},'' \href{https://doi.org/10.1109/MCG.2012.24}{\emph{IEEE
  Computer Graphics and Applications}},
  \href{https://doi.org/10.1109/MCG.2012.24}{vol.~32},
  \href{https://doi.org/10.1109/MCG.2012.24}{no.~1},
  \href{https://doi.org/10.1109/MCG.2012.24}{pp. 12--19},
  \href{https://doi.org/10.1109/MCG.2012.24}{Jan. 2012}.
  \href{https://doi.org/10.1109/MCG.2012.24}
{doi: {{%
10\hspace{.1pt}\discretionary{.}{%
}{.}\hspace{.4pt}1109\discretionary{/}{%
}{/}MCG\hspace{.1pt}\discretionary{.}{%
}{.}\hspace{.4pt}2012\hspace{.1pt}\discretionary{.}{%
}{.}\hspace{.4pt}24}}}


\bibitem{Madhyastha1995}
\href{https://doi.org/10.1109/52.368264}{T.~M. Madhyastha and D.~A. Reed},
  \href{https://doi.org/10.1109/52.368264}{``Data sonification: Do you see what
  {I} hear?}'' \href{https://doi.org/10.1109/52.368264}{\emph{IEEE Software}},
  \href{https://doi.org/10.1109/52.368264}{vol.~12},
  \href{https://doi.org/10.1109/52.368264}{no.~2},
  \href{https://doi.org/10.1109/52.368264}{p. 45–56},
  \href{https://doi.org/10.1109/52.368264}{Mar. 1995}.
  \href{https://doi.org/10.1109/52.368264}
{doi: {{%
10\hspace{.1pt}\discretionary{.}{%
}{.}\hspace{.4pt}1109\discretionary{/}{%
}{/}52\hspace{.1pt}\discretionary{.}{%
}{.}\hspace{.4pt}368264}}}


\bibitem{marion}
\href{https://doi.org/10.1109/TVCG.2017.2744518}{P.~Mindek, D.~Kou\v{r}il,
  J.~Sorger, D.~Toloudis, B.~Lyons, G.~Johnson, M.~E. Gr{\"o}ller, and
  I.~Viola}, \href{https://doi.org/10.1109/TVCG.2017.2744518}{``Visualization
  multi-pipeline for communicating biology},''
  \href{https://doi.org/10.1109/TVCG.2017.2744518}{\emph{IEEE Transactions on
  Visualization and Computer Graphics}},
  \href{https://doi.org/10.1109/TVCG.2017.2744518}{vol.~24},
  \href{https://doi.org/10.1109/TVCG.2017.2744518}{no.~1},
  \href{https://doi.org/10.1109/TVCG.2017.2744518}{pp. 883--892},
  \href{https://doi.org/10.1109/TVCG.2017.2744518}{Jan. 2017}.
  \href{https://doi.org/10.1109/TVCG.2017.2744518}
{doi: {{%
10\hspace{.1pt}\discretionary{.}{%
}{.}\hspace{.4pt}1109\discretionary{/}{%
}{/}TVCG\hspace{.1pt}\discretionary{.}{%
}{.}\hspace{.4pt}2017\hspace{.1pt}\discretionary{.}{%
}{.}\hspace{.4pt}2744518}}}


\bibitem{mindek2015}
\href{https://doi.org/10.1145/2788539.2788549}{P.~Mindek, L.~\v{C}mol\'{i}k,
  I.~Viola, M.~E. Gr{\"o}ller, and S.~Bruckner},
  \href{https://doi.org/10.1145/2788539.2788549}{``Automatized summarization of
  multiplayer games},'' \href{https://doi.org/10.1145/2788539.2788549}{in
  \emph{Proc.\ SCCG}}.\hskip 1em plus 0.5em minus 0.4em\relax
  \href{https://doi.org/10.1145/2788539.2788549}{Bratislava: Comenius
  University Bratislava, 2015},
  \href{https://doi.org/10.1145/2788539.2788549}{pp. 73--80}.
  \href{https://doi.org/10.1145/2788539.2788549}
{doi: {{%
10\hspace{.1pt}\discretionary{.}{%
}{.}\hspace{.4pt}1145\discretionary{/}{%
}{/}2788539\hspace{.1pt}\discretionary{.}{%
}{.}\hspace{.4pt}2788549}}}


\bibitem{cellview}
\href{https://doi.org/10.2312/vcbm.20151209}{M.~L. Muzic, L.~Autin, J.~Parulek,
  and I.~Viola}, \href{https://doi.org/10.2312/vcbm.20151209}{``{cellVIEW}: A
  tool for illustrative and multi-scale rendering of large biomolecular
  datasets},'' \href{https://doi.org/10.2312/vcbm.20151209}{in \emph{Proc.\
  VCBM}}.\hskip 1em plus 0.5em minus 0.4em\relax
  \href{https://doi.org/10.2312/vcbm.20151209}{Goslar, Germany: Eurographics
  Assoc., 2015}, \href{https://doi.org/10.2312/vcbm.20151209}{pp. 61--70}.
  \href{https://doi.org/10.2312/vcbm.20151209}
{doi: {{%
10\hspace{.1pt}\discretionary{.}{%
}{.}\hspace{.4pt}2312\discretionary{/}{%
}{/}vcbm\hspace{.1pt}\discretionary{.}{%
}{.}\hspace{.4pt}20151209}}}


\bibitem{nageli:2017:drone}
\href{https://doi.org/10.1145/3072959.3073712}{T.~N\"{a}geli, L.~Meier,
  A.~Domahidi, J.~Alonso-Mora, and O.~Hilliges},
  \href{https://doi.org/10.1145/3072959.3073712}{``Real-time planning for
  automated multi-view drone cinematography},''
  \href{https://doi.org/10.1145/3072959.3073712}{\emph{ACM Transactions on
  Graphics}}, \href{https://doi.org/10.1145/3072959.3073712}{vol.~36},
  \href{https://doi.org/10.1145/3072959.3073712}{no.~4},
  \href{https://doi.org/10.1145/3072959.3073712}{pp. 132:1--132:10},
  \href{https://doi.org/10.1145/3072959.3073712}{Jul. 2017}.
  \href{https://doi.org/10.1145/3072959.3073712}
{doi: {{%
10\hspace{.1pt}\discretionary{.}{%
}{.}\hspace{.4pt}1145\discretionary{/}{%
}{/}3072959\hspace{.1pt}\discretionary{.}{%
}{.}\hspace{.4pt}3073712}}}


\bibitem{nguyen2020modeling}
\href{https://arxiv.org/abs/2005.01804}{N.~Nguyen, O.~Strnad, T.~Klein, D.~Luo,
  R.~Alharbi, P.~Wonka, M.~Maritan, P.~Mindek, L.~Autin, D.~S. Goodsell, and
  I.~Viola}, \href{https://arxiv.org/abs/2005.01804}{``Modeling in the time of
  {COVID-19}: Statistical and rule-based mesoscale models},''
  \href{https://arxiv.org/abs/2005.01804}{arXiv preprint},
  \href{https://arxiv.org/abs/2005.01804}{2020}.

\bibitem{oskam:2009:visibility}
\href{https://doi.org/10.1145/1599470.1599478}{T.~Oskam, R.~W. Sumner,
  N.~Thuerey, and M.~Gross},
  \href{https://doi.org/10.1145/1599470.1599478}{``Visibility transition
  planning for dynamic camera control},''
  \href{https://doi.org/10.1145/1599470.1599478}{in \emph{Proc.\ SCA}}.\hskip
  1em plus 0.5em minus 0.4em\relax
  \href{https://doi.org/10.1145/1599470.1599478}{New York: ACM, 2009},
  \href{https://doi.org/10.1145/1599470.1599478}{p. 55–65}.
  \href{https://doi.org/10.1145/1599470.1599478}
{doi: {{%
10\hspace{.1pt}\discretionary{.}{%
}{.}\hspace{.4pt}1145\discretionary{/}{%
}{/}1599470\hspace{.1pt}\discretionary{.}{%
}{.}\hspace{.4pt}1599478}}}


\bibitem{ren2017}
\href{https://doi.org/10.1109/PACIFICVIS.2017.8031599}{D.~Ren, M.~Brehmer,
  B.~Lee, T.~H{\"o}llerer, and E.~K. Choe},
  \href{https://doi.org/10.1109/PACIFICVIS.2017.8031599}{``Chart{A}ccent:
  Annotation for data-driven storytelling},''
  \href{https://doi.org/10.1109/PACIFICVIS.2017.8031599}{in \emph{Proc.\
  PacificVis}}.\hskip 1em plus 0.5em minus 0.4em\relax
  \href{https://doi.org/10.1109/PACIFICVIS.2017.8031599}{Los Alamitos: IEEE
  Computer Society, 2017},
  \href{https://doi.org/10.1109/PACIFICVIS.2017.8031599}{pp. 230--239}.
  \href{https://doi.org/10.1109/PACIFICVIS.2017.8031599}
{doi: {{%
10\hspace{.1pt}\discretionary{.}{%
}{.}\hspace{.4pt}1109\discretionary{/}{%
}{/}PACIFICVIS\hspace{.1pt}\discretionary{.}{%
}{.}\hspace{.4pt}2017\hspace{.1pt}\discretionary{.}{%
}{.}\hspace{.4pt}8031599}}}


\bibitem{salomon:2003:interactive}
\href{https://doi.org/10.1145/641480.641491}{B.~Salomon, M.~Garber, M.~C. Lin,
  and D.~Manocha}, \href{https://doi.org/10.1145/641480.641491}{``Interactive
  navigation in complex environments using path planning},''
  \href{https://doi.org/10.1145/641480.641491}{in \emph{Proc.\ I3D}}.\hskip 1em
  plus 0.5em minus 0.4em\relax \href{https://doi.org/10.1145/641480.641491}{New
  York: ACM, 2003}, \href{https://doi.org/10.1145/641480.641491}{pp. 41--50}.
  \href{https://doi.org/10.1145/641480.641491}
{doi: {{%
10\hspace{.1pt}\discretionary{.}{%
}{.}\hspace{.4pt}1145\discretionary{/}{%
}{/}641480\hspace{.1pt}\discretionary{.}{%
}{.}\hspace{.4pt}641491}}}


\bibitem{segel-heer:2010:narrative-visualization}
\href{https://doi.org/10.1109/TVCG.2010.179}{E.~{Segel} and J.~{Heer}},
  \href{https://doi.org/10.1109/TVCG.2010.179}{``Narrative visualization:
  Telling stories with data},''
  \href{https://doi.org/10.1109/TVCG.2010.179}{\emph{IEEE Transactions on
  Visualization and Computer Graphics}},
  \href{https://doi.org/10.1109/TVCG.2010.179}{vol.~16},
  \href{https://doi.org/10.1109/TVCG.2010.179}{no.~6},
  \href{https://doi.org/10.1109/TVCG.2010.179}{pp. 1139--1148},
  \href{https://doi.org/10.1109/TVCG.2010.179}{Nov 2010}.
  \href{https://doi.org/10.1109/TVCG.2010.179}
{doi: {{%
10\hspace{.1pt}\discretionary{.}{%
}{.}\hspace{.4pt}1109\discretionary{/}{%
}{/}TVCG\hspace{.1pt}\discretionary{.}{%
}{.}\hspace{.4pt}2010\hspace{.1pt}\discretionary{.}{%
}{.}\hspace{.4pt}179}}}


\bibitem{seligmann:1993:supporting}
\href{https://doi.org/10.1145/169891.169896}{D.~D. Seligmann and S.~Feiner},
  \href{https://doi.org/10.1145/169891.169896}{``Supporting interactivity in
  automated {3D} illustrations},''
  \href{https://doi.org/10.1145/169891.169896}{in \emph{Proc.\ IUI}}.\hskip 1em
  plus 0.5em minus 0.4em\relax \href{https://doi.org/10.1145/169891.169896}{New
  York: ACM, 1993}, \href{https://doi.org/10.1145/169891.169896}{pp. 37--44}.
  \href{https://doi.org/10.1145/169891.169896}
{doi: {{%
10\hspace{.1pt}\discretionary{.}{%
}{.}\hspace{.4pt}1145\discretionary{/}{%
}{/}169891\hspace{.1pt}\discretionary{.}{%
}{.}\hspace{.4pt}169896}}}


\bibitem{Siddhi2017}
\href{https://doi.org/10.5120/ijca2017913891}{D.~Siddhi, J.~M. Verghese, and
  D.~Bhavik}, \href{https://doi.org/10.5120/ijca2017913891}{``Survey on various
  methods of text to speech synthesis},''
  \href{https://doi.org/10.5120/ijca2017913891}{\emph{International Journal of
  Computer Applications}}, \href{https://doi.org/10.5120/ijca2017913891}{vol.
  165}, \href{https://doi.org/10.5120/ijca2017913891}{no.~6},
  \href{https://doi.org/10.5120/ijca2017913891}{pp. 26--30},
  \href{https://doi.org/10.5120/ijca2017913891}{May 2017}.
  \href{https://doi.org/10.5120/ijca2017913891}
{doi: {{%
10\hspace{.1pt}\discretionary{.}{%
}{.}\hspace{.4pt}5120\discretionary{/}{%
}{/}ijca2017913891}}}


\bibitem{sorger2017}
\href{https://doi.org/10.1145/3154353.3154364}{J.~Sorger, P.~Mindek, P.~Rautek,
  M.~E. Gr{\"{o}}ller, G.~Johnson, and I.~Viola},
  \href{https://doi.org/10.1145/3154353.3154364}{``Metamorphers: Storytelling
  templates for illustrative animated transitions in molecular
  visualization},'' \href{https://doi.org/10.1145/3154353.3154364}{in
  \emph{Proc.\ SCCG}}.\hskip 1em plus 0.5em minus 0.4em\relax
  \href{https://doi.org/10.1145/3154353.3154364}{New York: ACM, 2017},
  \href{https://doi.org/10.1145/3154353.3154364}{pp. 27--36}.
  \href{https://doi.org/10.1145/3154353.3154364}
{doi: {{%
10\hspace{.1pt}\discretionary{.}{%
}{.}\hspace{.4pt}1145\discretionary{/}{%
}{/}3154353\hspace{.1pt}\discretionary{.}{%
}{.}\hspace{.4pt}3154364}}}


\bibitem{thony2018}
\href{https://doi.org/10.3390/ijgi7030123}{M.~Th{\"o}ny, R.~Schn{\"u}rer,
  R.~Sieber, L.~Hurni, and R.~Pajarola},
  \href{https://doi.org/10.3390/ijgi7030123}{``Storytelling in interactive {3D}
  geographic visualization systems},''
  \href{https://doi.org/10.3390/ijgi7030123}{\emph{ISPRS International Journal
  of Geo-Information}}, \href{https://doi.org/10.3390/ijgi7030123}{vol.~7},
  \href{https://doi.org/10.3390/ijgi7030123}{no.~3},
  \href{https://doi.org/10.3390/ijgi7030123}{pp. 123:1--123:14},
  \href{https://doi.org/10.3390/ijgi7030123}{Mar. 2018}.
  \href{https://doi.org/10.3390/ijgi7030123}
{doi: {{%
10\hspace{.1pt}\discretionary{.}{%
}{.}\hspace{.4pt}3390\discretionary{/}{%
}{/}ijgi7030123}}}


\bibitem{tong:2018:storytelling}
\href{https://doi.org/10.3390/info9030065}{C.~Tong, R.~Roberts, R.~Borgo,
  S.~Walton, R.~S. Laramee, K.~Wegba, A.~Lu, Y.~Wang, H.~Qu, Q.~Luo, and
  X.~Ma}, \href{https://doi.org/10.3390/info9030065}{``Storytelling and
  visualization: An extended survey},''
  \href{https://doi.org/10.3390/info9030065}{\emph{Information}},
  \href{https://doi.org/10.3390/info9030065}{vol.~9},
  \href{https://doi.org/10.3390/info9030065}{no.~3},
  \href{https://doi.org/10.3390/info9030065}{pp. 65:1--65:42},
  \href{https://doi.org/10.3390/info9030065}{Mar. 2018}.
  \href{https://doi.org/10.3390/info9030065}
{doi: {{%
10\hspace{.1pt}\discretionary{.}{%
}{.}\hspace{.4pt}3390\discretionary{/}{%
}{/}info9030065}}}


\bibitem{vanWijk:2003:zooming-and-panning}
\href{https://doi.org/10.1109/INFVIS.2003.1249004}{J.~J. {van Wijk} and
  W.~A.~A. {Nuij}}, \href{https://doi.org/10.1109/INFVIS.2003.1249004}{``Smooth
  and efficient zooming and panning},''
  \href{https://doi.org/10.1109/INFVIS.2003.1249004}{in \emph{Proc.\
  InfoVis}}.\hskip 1em plus 0.5em minus 0.4em\relax
  \href{https://doi.org/10.1109/INFVIS.2003.1249004}{Los Alamitos: IEEE
  Computer Society, 2003},
  \href{https://doi.org/10.1109/INFVIS.2003.1249004}{pp. 15--23}.
  \href{https://doi.org/10.1109/INFVIS.2003.1249004}
{doi: {{%
10\hspace{.1pt}\discretionary{.}{%
}{.}\hspace{.4pt}1109\discretionary{/}{%
}{/}INFVIS\hspace{.1pt}\discretionary{.}{%
}{.}\hspace{.4pt}2003\hspace{.1pt}\discretionary{.}{%
}{.}\hspace{.4pt}1249004}}}


\bibitem{Varner:2014:biology-science-outreach}
\href{https://doi.org/10.1093/biosci/biu021}{J.~Varner},
  \href{https://doi.org/10.1093/biosci/biu021}{``Scientific outreach: Toward
  effective public engagement with biological science},''
  \href{https://doi.org/10.1093/biosci/biu021}{\emph{BioScience}},
  \href{https://doi.org/10.1093/biosci/biu021}{vol.~64},
  \href{https://doi.org/10.1093/biosci/biu021}{no.~4},
  \href{https://doi.org/10.1093/biosci/biu021}{pp. 333--340},
  \href{https://doi.org/10.1093/biosci/biu021}{Mar. 2014}.
  \href{https://doi.org/10.1093/biosci/biu021}
{doi: {{%
10\hspace{.1pt}\discretionary{.}{%
}{.}\hspace{.4pt}1093\discretionary{/}{%
}{/}biosci\discretionary{/}{%
}{/}biu021}}}


\bibitem{Vazquez:2008:interactive-anatom-edu}
\href{https://doi.org/10.1007/s11548-008-0251-4}{P.-P. V{\'a}zquez,
  T.~G{\"o}tzelmann, K.~Hartmann, and A.~N{\"u}rnberger},
  \href{https://doi.org/10.1007/s11548-008-0251-4}{``An interactive {3D}
  framework for anatomical education},''
  \href{https://doi.org/10.1007/s11548-008-0251-4}{\emph{International Journal
  of Computer Assisted Radiology and Surgery}},
  \href{https://doi.org/10.1007/s11548-008-0251-4}{vol.~3},
  \href{https://doi.org/10.1007/s11548-008-0251-4}{no.~6},
  \href{https://doi.org/10.1007/s11548-008-0251-4}{pp. 511--524},
  \href{https://doi.org/10.1007/s11548-008-0251-4}{Aug. 2008}.
  \href{https://doi.org/10.1007/s11548-008-0251-4}
{doi: {{%
10\hspace{.1pt}\discretionary{.}{%
}{.}\hspace{.4pt}1007\discretionary{/}{%
}{/}s11548\discretionary{%
}{-}{-}008\discretionary{%
}{-}{-}0251\discretionary{%
}{-}{-}4}}}


\bibitem{Wilson2020}
\href{https://doi.org/1853/50809}{C.~M. Wilson and S.~K. Lodha},
  \href{https://doi.org/1853/50809}{``Listen: A data sonification toolkit},''
  \href{https://doi.org/1853/50809}{in \emph{Proc.\ ICAD}}.\hskip 1em plus
  0.5em minus 0.4em\relax \href{https://doi.org/1853/50809}{Atlanta: Georgia
  Institute of Technology, 1996}. \href{https://doi.org/1853/50809}
{doi: {{%
1853\discretionary{/}{%
}{/}50809}}}


\bibitem{Wohlfart:2007}
\href{https://doi.org/10.2312/VisSym/EuroVis07/091-098}{M.~Wohlfart and
  H.~Hauser}, \href{https://doi.org/10.2312/VisSym/EuroVis07/091-098}{``Story
  telling for presentation in volume visualization},''
  \href{https://doi.org/10.2312/VisSym/EuroVis07/091-098}{in \emph{Proc.\
  VisSym}}.\hskip 1em plus 0.5em minus 0.4em\relax
  \href{https://doi.org/10.2312/VisSym/EuroVis07/091-098}{Goslar, Germany:
  Eurographics Assoc., 2007},
  \href{https://doi.org/10.2312/VisSym/EuroVis07/091-098}{pp. 91--98}.
  \href{https://doi.org/10.2312/VisSym/EuroVis07/091-098}
{doi: {{%
10\hspace{.1pt}\discretionary{.}{%
}{.}\hspace{.4pt}2312\discretionary{/}{%
}{/}VisSym\discretionary{/}{%
}{/}EuroVis07\discretionary{/}{%
}{/}091\discretionary{%
}{-}{-}098}}}


\end{thebibliography}
\end{document}